\documentclass[aps,pre,onecolumn, amsfonts,showpacs, superscriptaddress]{revtex4-1}
\usepackage{epsfig,amsopn}
\usepackage{graphicx}
\usepackage{amsmath,amssymb}
\usepackage{braket}
\usepackage{color}
\newcommand{\lmax}{\lambda_{\max}}

\newcommand{\e}{{\rm e}}

\def\bea{\begin{eqnarray}}
\def\eea{\end{eqnarray}}

\def\nn{\nonumber}

\def\f{\frac}

\begin{document}

\title{Extreme value statistics of correlated random variables: a pedagogical review}

\author{Satya N. \surname{Majumdar}}
\affiliation{LPTMS, CNRS, Univ. Paris-Sud, Universit\'e Paris-Saclay, 91405 Orsay, France}

\author{Arnab Pal} \affiliation{School of Chemistry, The Center for Physics and Chemistry of Living Systems,
The Raymond and Beverly Sackler Center for Computational Molecular and Materials Science,
\& The Mark Ratner Institute for Single Molecule Chemistry, Tel Aviv University, Tel Aviv 6997801, Israel}

\author{Gr\'egory \surname{Schehr}}
\affiliation{LPTMS, CNRS, Univ. Paris-Sud, Universit\'e Paris-Saclay, 91405 Orsay, France}

\date{\today}

\begin{abstract} Extreme value statistics (EVS) concerns the study of the statistics of the maximum or the minimum 
of a set of random variables. This is an important problem for any time-series and has applications in climate, 
finance, sports, all the way to physics of disordered systems where one is interested in the statistics of the 
ground state energy. While the EVS of `uncorrelated' variables are well understood, little is known for strongly 
correlated random variables. Only recently this subject has gained much importance both in statistical physics and 
in probability theory. In this review, we will first recall the classical EVS for uncorrelated variables and discuss the
three universality classes of extreme value limiting distribution, known as the Gumbel, Fr\'echet and
Weibull distribution. We then show that,
for weakly correlated random variables with a finite correlation length/time, the limiting extreme value distribution
can still be inferred from that of the uncorrelated variables using a renormalisation group-like argument. Finally, we consider 
the most interesting examples of strongly correlated variables for which there are very few exact results for the EVS. We discuss
few examples of such strongly correlated systems (such as the Brownian motion and the eigenvalues of
a random matrix) where some analytical progress can be made. We also discuss other observables related to extremes, such as the density of near-extreme events, time at which an extreme value occurs, order and record statistics, etc. 
\end{abstract}

\maketitle

\tableofcontents

\section{Introduction}
\label{introduction}

Extreme events are ubiquitous in nature. They may be rare events but when they
occur, they may have devastating consequences and hence are rather
important from practical points of view. 
To name a few, different forms of natural calamities
such as earthquakes, tsunamis, extreme floods, large wildfire, the hottest and the coldest days,
stock market risks or 
large insurance losses in finance, new records in major sports events like Olympics 
are typical examples of 
extreme events. There has been a major interest to study these 
events systematically using statistical methods and the field is known as Extreme Value Statistics
(EVS)~\cite{FT28,Gumbel,Gnedenko,Leadbetter}. This is a branch of statistics dealing with the extreme deviations from the mean/median 
of probability distributions. The general theory sets out to assess the type of 
probability distributions generated by processes which are responsible for
these kinds of highly unusual events. In recent years, it has been realized
that extreme statistics and rare events play an equally important role
in various physical, biological and financial contexts as well--for a few illustrative examples (by far not exhaustive) 
see~\cite{Derrida:81,Derrida:85, DG86, TW94, TW96,Bouchaud:97,KM00,Satya:00,Dean:01,ADGR01,Raychowdhury,Satya:02,Racz:03,Ledoussal-Monthus,
MK03,Satya:04,airy,KM05,Satya:05,Bertin-Clusel,GMOR07,Sanjib,BeM07,Krug07,RM07,CS07,CS107,BGMZ07,Burkhardt,Satya:08,MB08,
MRKY08,MZ08,GL08,KIK08,SMCR08,KIK08b,Fei09,RMC09,LW09,GMS09,MCR10,CBM15,MRZ10,MRZ110,SL10,RS11,NK11,RS11b,FQR13,FMS11,Lie12,S12,SM13,SMCF13,BLS12,Wergen13,DMRZ13,GMS14,FC15,GMS17,SHMS18,BM07,FB07,FB08,Satya_Ziv,BenNaim}.
A typical example 
can be found in disordered systems where the ground state energy, being
the minimum energy, plays the role of an extreme variable. In addition,
the dynamics at low temperatures in disordered systems are governed by the 
statistics of the highest energy barrier in the system. Hence the study of extremes and related quantities
is extremely important in the field of disordered systems~\cite{Derrida:81,Derrida:85,DG86,Bouchaud:97,Satya:00,Dean:01,Ledoussal-Monthus,FB08,FB08a,F09,FLR09,F10,MRZ10}. 
A rather important physical system where extreme fluctuations play
an important role corresponds to fluctuating interfaces in the
{Edwards-Wilkinson or Kardar-Parisi-Zhang universality classes}~\cite{Raychowdhury,Racz:03,Satya:04,airy,SM06,GMOR07,Burkhardt,RS09}.
Another exciting recent area concerns the distribution of the
largest eigenvalue in random matrices: the limiting distribution~\cite{TW94,TW96} and 
the large deviation probabilities~\cite{DM06,DM08,MV09} of the
largest eigenvalue and its various
applications (for a recent review on the largest eigenvalue of a random matrix, see~\cite{Satya:14}). 
Extreme value statistics also
appears in computer science problems such as in binary search trees
and related search algorithms~\cite{KM00,Satya:00,Satya:02,MK03,Satya:05}.

In the classical extreme value theory, one is concerned with the statistics
of the maximum (or minimum) of a set of {\em uncorrelated} random variables {(see e.g. the classical textbooks \cite{Gumbel, Leadbetter} or the review \cite{FC15})}.
In contrast, in most of the physical systems mentioned above, the
underlying random variables are typically {\em correlated}. 
In recent years, there have been some advances in the understanding
of EVS of correlated variables. In this survey, we will first review
the classical EVS of uncorrelated variables. Then we will discuss 
the EVS of {\em weakly} correlated random variables with some examples.
Finally few examples of {\em strongly} correlated random variables will
be discussed. 

This subject of EVS is rapidly evolving and this is a short review on some of the recent
developments and the key-questions in the field. The purpose of this review is not to provide 
detailed calculations for different models, but rather point out to the reader the relevant questions, 
recent developments and the relevant references. The interested reader can look up this literature
for details. The choice of topics reflects our personal taste
and contributions to the field and, by no means, is exhaustive. Hence, the list of references is clearly
far from being exhaustive and any inadvertent omission of a relevant reference is apologized.

\section{Extreme value statistics: basic preliminaries}

In a given physical situation, one needs to first identify
the set of relevant random variables $\{x_1,x_2,\ldots, x_N\}$. For example, for fluctuating
one-dimensional interfaces, the relevant random variables may denote
the heights of 
the interface at different space points. In disordered systems
such as spin glasses, $\{x_i\}$'s may denote the energy of different
spin configurations for a given sample of quenched disorder. Once
the random variables are identified, there are subsequently two basic steps
involved :
(i) to compute explicitly the joint distribution
$P(\{x_i\})$ of the relevant random variables (this is sometimes
very difficult to achieve) and {(ii) from the joint
distribution $P(\{x_i\})$ calculate
the distribution of some observables, such as the sample mean
or the sample maximum, defined as}
\bea
\text{mean}~~~\bar{X}&=&\f{x_{1}+x_{2}+...+x_{N}}{N}~, \label{mean}\\
\text{maximum}~~M&=&\text{max}~(x_{1},x_{2},...,x_{N})\,. \label{max}
\eea

Particular simplifications occur for independent and identically distributed (IID)
random variables, where the
joint distribution $P(\{x_i\})$ factorizes, i.e., 
$P(x_{1},x_{2},...,x_{N})=p(x_{1})p(x_{2})...p(x_{N})$, where
each variable is chosen from the same parent probability density 
$p(x)$. Knowing the parent distribution $p(x)$, one can then easily
compute the distributions, e.g., of $\bar{X}$ and of $M$. 

For example, let us first consider $\bar{X}$. 
One knows that irrespective of the choice of the parent distribution (with finite variance)
the probability distribution function (PDF) of the {\it mean} 
of $N$ IID random variables tends to a Gaussian distribution for large $N$ namely,
\bea
P(\bar{X},N)\xrightarrow{N\to \infty}\f{1}{\sqrt{2\pi \sigma^{2}/N}}\e^{-\f{N}{2 \sigma^{2}}(\bar{X}-\mu)^{2}} \;,
\eea
where $\mu$ and $\sigma^{2}$ are the mean and the variance of the parent distribution respectively. This is 
known as the {\it Central Limit Theorem} \cite{Feller:71} and this Gaussian form is universal. 
However, for correlated variables, one does not know, in general, how to compute
the distribution of $\bar{X}$ {and a priory one expects that the limiting distribution of $\bar{X}$
will be different from a Gaussian (see for instance \cite{Bertin-Clusel} for a discussion of this question in connection with extreme statistics).}

Similar question about universality also arises for the distribution of extremes, e.g.,
that of $M$ in Eq. (\ref{max}). We will see below that, as in the case of the mean $\bar{X}$, there
exist universal limit laws for the distribution of the maximum $M$
for the case of IID variables. However, for {\em strongly} correlated
variables, the issue of universality is wide open. Suppose that we know the joint distribution $P(\{x_i\})$ explicitly.
Then to compute the distribution of the maximum $M$, it is useful to
define the cumulative distribution of $M$ which can be easily expressed
in terms of the joint distribution
\bea \label{def_cumul}
Q_{N}(x)&=&\text{Prob}[M\leq x,N]=\text{Prob}[x_{1}\leq x,~x_{2}\leq x,...,x_{N}\leq x]~,\\
&=&\int_{-\infty}^{x} dx_1
\int_{-\infty}^{x} dx_2 \cdots \int_{-\infty}^{x}dx_{N} P(x_{1},x_{2}, \cdots ,x_{N})~,
\eea
and the PDF of the maximum can be obtained by taking the derivative i.e. $P(M,N)=Q_{N}^{\prime}(M)\equiv \frac{dQ_N}{dx} \large |_{x=M}$.

\section{Independent and Identically distributed random variables}\label{sec:IID}

For IID random variables, the joint PDF factorizes and we get
\bea\label{def_QN}
Q_{N}(x)=\left[\int_{-\infty}^{x}dy~p(y)\right]^{N}=\left[1-\int_{x}^{\infty}dy~p(y)\right]^{N}\,.
\eea 
This is an exact formula for the cumulative distribution of the maximum for any $N$.
Evidently, $Q_N(x)$ depends explicitly on the parent distribution $p(x)$ for any
finite $N$. The question is: as in the CLT of the sum of random variables discussed
before, does any universality emerge for $Q_N(x)$ in the large $N$ limit? 
The answer is that indeed a form of universality emerges in the large $N$ limit, as
we summarize below.

It turns out that in the scaling limit when $N$ is large, $x$ is large, with
a particular scaling combination (see below) fixed, $Q_N(x)$ approaches a limiting form:
\bea
Q_{N}(x)\xrightarrow[z=(x-a_{N})/b_{N}~\text{fixed}]{x \to \infty,~N\to \infty}F\left(\f{x-a_{N}}{b_{N}}\right) \label{scaling_iid}\\
\text{equivalently}~,~~~\lim_{N \to \infty}Q_{N}(a_{N}+b_{N}z)=F(z)
\eea
where $a_{N},~b_{N}$ are non-universal scaling factors that depend on the parent distribution $p(x)$,
but the scaling function $F(z)$ can only be of the three possible forms
$F_{1,2,3}(z)$ depending only on the large $x$ tail of the parent distribution
$p(x)$.   
This is known as the Gnedenko's classical law of extremes~\cite{Gnedenko}.

\subsection{Parent distributions with a power law tail}

We consider the IID random variables whose parent distribution has a power law convergence 
$p(x)\sim A\,x^{-(1+\alpha)}$ with $A, \alpha>0$. In this case, we denote the scaling function as $F_{1}(z)$ 
and this is found to be 
\bea\label{F1}
F_{1}(z)= \left\{
     \begin{array} {rl}
     {\rm e}^{-z^{-\alpha}} & ~ \text{for~} z \geq 0\\
     {\rm 0}  & ~ \text{for~}  z \leq 0 \;. \\
     \end{array} \right.
\eea
The PDF is given by 
\bea
f_{1}(z)=F_1'(z) = \f{\alpha}{z^{\alpha+1}}{\rm e}^{-z^{-\alpha}},~~~~~z\in[0,\infty)~.
\eea
Here one can identify $a_{N}=0,~b_{N}=(A\,N/\alpha)^{1/\alpha}$. This is the famous {\it Fr\'{e}chet} distribution.

\subsection{Parent distributions with a faster than power law tail}

We consider the parent distributions with tails that decay faster than a power law, but are still unbounded,
such as $p(x)\sim \e^{-x^{\delta}}$ with $\delta>0$. In this case, one finds the scaling function to be
\bea
F_{2}(z)=\e^{-\e^{-z}} \label{F2}~,
\eea
where the PDF is given by
\bea
f_{2}(z)=F_2'(z) = \e^{-z-\e^{-z}},~~~~~z\in(-\infty,\infty) \;. \label{def_Gumbel}
\eea
{This is the famous {\it Gumbel} distribution~\cite{FT28,Gumbel}. Here one finds
$a_{N}=(\ln N)^{1/\delta},~b_{N}=\f{1}{\delta}(\ln N)^{1/\delta-1}$.
Since $a_{N}$ is the typical value of the maximum and is defined by
\begin{eqnarray}\label{def_aN}
N\,\int^\infty_{a_{N}}p(y)dy=1~,
\end{eqnarray}
we see from Eq. (\ref{scaling_iid}) that the weight will be to the right of $a_{N}$, since $F_2(z)$ in Eq. (\ref{F2}) decays extremely rapidly to $0$ for negative argument.}
In the following we will
compute the scaling functions for the parent distributions having 
the exponential and the Gaussian tails.

\subsubsection{The exponential case $p(x) = \e^{-x}$ {\rm for}\, $x \geq 0$}
In this case, applying the form of $p(x)$, one finds
\bea
Q_{N}(x)=[1-\e^{-x}]^{N}=\e^{N\log[1-\e^{-x}]}\sim \e^{-N \e^{-x}}=\e^{-\e^{-(x-\log N)}}=F_{2}(z)
\eea 
with $z=x-\log N$ and the following identification $a_{N}=\ln N,~b_{N}=1$ in Eq. (\ref{scaling_iid}).
\subsubsection{The Gaussian case $p(x) = \frac{1}{\sqrt{2\pi}} \e^{-x^{2}/2}$}

As in the previous example, substituting this form of $p(x)$ in Eq.~(\ref{def_QN}) we obtain 
\bea\label{Cumul_Gaussian1}
Q_N(x) = \left[ 1 - \frac{1}{2}\, {\rm erfc}\left(\frac{x}{\sqrt{2}}\right)\right]^N  = \exp{\left[ N \ln\left(1 -\frac{1}{2}\, {\rm erfc}\left(\frac{x}{\sqrt{2}}\right) \right)\right]} ~,
\eea 
where ${\rm erfc}(x) = \frac{2}{\sqrt{\pi}} \int_x^{\infty} \e^{-y^2} \, dy$ is the complementary error function. Expanding the logarithm in Taylor series for large $x$, we get an approximate formula
\bea\label{Cumul_Gaussian2}
Q_N(x) \approx \exp{\left[-\frac{N}{2} {\rm erfc}\left(\frac{x}{\sqrt{2}} \right)  \right]} \;.
\eea
We only want to retain the leading behaviour for large $x$. This can be done by using the large $x$ behaviour of ${\rm erfc}(x)$ which has the following form
\bea\label{erfc_large}
{\rm erfc}(x) \approx \frac{\e^{-x^2}}{x\,\sqrt{\pi}} \;, \; x \to \infty \;.
\eea
Finally, using this asymptotic behaviour in Eq. (\ref{Cumul_Gaussian2}), we get, for both $N$ and $x$ large (note that we still have not taken the scaling limit yet) the leading behaviour
\bea\label{Cumul_Gaussian3}
Q_N(x) \approx \exp{\left( - \frac{N}{\sqrt{2 \pi}} \frac{\e^{-x^2/2}}{x} \right)} \;.
\eea
To cast this tail behavior in the universal form as in Eq. (\ref{scaling_iid}), we look for $a_N$ and $b_N$ such that $(x-a_N)/b_N = z$ is fixed as $N \to \infty$. For this, we set $x = a_N + b_N \, z$ in Eq.~(\ref{Cumul_Gaussian3}) to get
\bea\label{Cumul_Gaussian4}
Q_N(x=a_N + b_N \, z) \approx \exp{\left[ - \frac{N}{\sqrt{2 \pi}} \frac{1}{(a_N + b_N \, z)} \e^{-\frac{1}{2}(a_N^2 + 2 a_N \, b_N z + b_N^2 z^2)} \right]} \;.
\eea 
Let us anticipate (and verify a posteriori) that $b_N \ll a_N$ such that $(a_N + b_N \,z) \approx a_N$ to leading order and furthermore we can drop the term $b_N^2\,z^2$ inside the exponential, since we expect it to be small. We then choose $a_N$ and $b_N$ such that
\bea\label{aNbNGauss}
\frac{N}{\sqrt{2 \pi} a_N} \e^{-\frac{a_N^2}{2}} = 1 \;, \;\; {\rm and} \;\; a_N \, b_N = 1 \;.
\eea
With this choice of $a_N$ and $b_N$ we then have
\bea\label{Cumul_Gaussian5}
Q_N(x=a_N + b_N \, z) \approx \e^{-\e^{-z}} = F_2(z) \;.
\eea
The centring and the scaling constants $a_N$ and $b_N$ can be obtained for large $N$ from Eq. (\ref{aNbNGauss}). To leading order, one gets
\bea\label{aNbNGauss2}
a_N \approx \sqrt{2 \ln N} - \frac{\ln(\ln N)}{2 \sqrt{2 \ln N}} + \cdots \; \; {\rm and} \;\, b_N = \frac{1}{a_N} \;.
\eea
Thus we see that, even though the finite $N$ formula of $Q_N(x)$ in Eq. (\ref{def_QN}) for the two distributions $p(x) = \e^{-x}$ (with $x \geq 0$) and  $p(x) = \frac{1}{\sqrt{2 \pi}}\e^{-x^2/2}$ look rather different from each other at first glance, once we center and scale by appropriate $a_N$ and $b_N$ which differ in the two cases, the scaling function for the two cases is exactly identical and is given by the Gumbel law $F_2(z) = \e^{-\e^{-z}}$. 

One can now generalize the Gaussian case to any distribution with a large $x$ tail $p(x) \sim C\, \e^{-x^\delta}$ with $C, \delta > 0$. In this case, one can carry out a straightforward large $N$ analysis of the formula in Eq. (\ref{def_QN}), following the same method as discussed above. The centring factor $a_N$ and the width $b_N$ in Eq. (\ref{scaling_iid}) are given, for large $N$, by
\begin{eqnarray}\label{ab_delta}
a_N \approx (\ln N)^{1/\delta} \quad, \quad b_N \approx \frac{1}{\delta} (\ln N)^{1/\delta-1} \;.
\end{eqnarray}
The limiting scaling function however remains the Gumbel form $F_2(z) = \e^{-\e^{-z}}$ for any $\delta>0$. Interestingly, the width $b_N$ of the maximum distribution around its peak at $a_N$ increases for large $N$ for $0 < \delta <1$, while it decreases for large $N$ for $\delta > 1$. Thus, for $0 < \delta <1$, the PDF of the maximum becomes broader and broader with increasing $N$, while, for $\delta > 1$, the PDF becomes narrower and narrower around its peak.

\subsection{Parent distributions with an upper bounded support}

We now consider the parent distributions with bounded tails
such as $p(x)\xrightarrow{x \to a} (a-x)^{\beta-1}$ with $\beta>0$. 
In this case 
\bea \label{F3}
F_{3}(z)= \left\{
     \begin{array} {rl}
     { \e^{-(-z)^{\beta}}} & ~ \text{for~} z \leq 0\\
     {\rm 1}  & ~ \text{for~}  z \geq 0~. \\
     \end{array} \right.
\eea
The PDF is therefore given by
\bea
f_{3}(z)=\beta(-z)^{\beta-1}\e^{-(-z)^{\beta}},~~~~~z\in(-\infty,0]~,
\eea
which is the well-known {\it Weibull} distribution. Note that in the case where $p(x)$ vanishes faster than a power law at the upper bound, for instance if it has an essential singularity $p(x) \sim e^{-1/(a-x)^\nu}$ with $\nu>0$, then the limiting distribution of the maximum is given by a Gumbel law (\ref{F2}). 

Let us now summarize the results for the limiting distribution of the maximum
for IID random variables in the following table.

\begin{center}
    \begin{tabular}{ | l | l | l | p{3cm} |}
    \hline
    Parent distribution $p(x)$ & Scaling function $F(z)$ & PDF of maximum $f(z)$ & Nomenclature\\ \hline
    $x^{-(1+\alpha)};\,\, \alpha>0$ & $\e^{-z^{-\alpha}}\theta(z)$    
& $\f{\alpha}{z^{\alpha+1}}\e^{-z^{-\alpha}};\,\, z>0 $ &  \textbf{Fr\'{e}chet}  \\ \hline
    $\e^{-x^{\delta}}$ & $\e^{-\e^{-z}}$                  
& $\e^{-z-\e^{-z}}$                            &  \textbf{Gumbel}       \\ \hline
    $(a-x)^{\beta-1};\,\, \beta>0$ & $\e^{-(-z)^{\beta}}
\theta(-z)+\theta(z)$  & $\beta(-z)^{\beta-1}\e^{-(-z)^{\beta}};\,\,\, z<0$ &   \textbf{Weibull} \\
    \hline
    \end{tabular}
\end{center}

So far, we discussed about the limiting laws in the limit of large sample size $N$.
It turns out however that the convergence to these limiting laws is extremely
slow and in simulations and experiments, it is very hard to see these limiting distributions~\cite{GMOR08,astro}.
A renormalization group treatment has been developed that describes
how these EVS distributions for IID variables approach their limiting ``fixed
point'' distributions. Interested readers may consult Refs.~\cite{GMOR08,GMORD10,BG10,ABA12,CCEF12}.

In the discussion above, we have only considered the limiting large $N$ distributions that describe the {\it typical} 
fluctuations of the maximum value around its mean, i.e., when $x \sim a_N + b_N \,z$ where $z = {\cal O}(1)$. 
However, {\it atypical} fluctuations much larger than this typical scale, i.e., when $|z| \gg 1$, are not described by these 
limiting distributions, but rather by appropriate large deviation tails (for a recent discussion see \cite{PVivo}).

\subsection{Order statistics}

An interesting generalization of the statistics of the global maximum corresponds to studying
the statistics of successive maxima, known as the `order' statistics~\cite{ABN,ND}.
For a recent review on order statistics, see~\cite{SM13}. Consider again the set of IID random variables
$\{x_{1},x_{2},...,x_{N}\}$ and arrange them in decreasing
order of their values.
So, if we denote them by $M_{k,N}$ where $k$ is the order and $N$ is the number 
of variables, then
\bea \label{def_order}
M_{1,N}&=&\text{max~}(x_{1},x_{2}, \cdots ,x_{N})\;,  \nn \\
M_{2,N}&=&\text{second \,max~}(x_{1},x_{2}, \cdots ,x_{N})\;, \nn \\
\vdots \nn \\
M_{k,N}&=&\text{$k$-th \,max~}(x_{1},x_{2}, \cdots ,x_{N})\;,  \nn \\
\vdots \nn \\
M_{N,N}&=&\text{min~}(x_{1},x_{2}, \cdots ,x_{N}) \;,
\eea
and henceforth by definition 
$M_{1,N}>M_{2,N}> \ldots >M_{N,N}$. It is of interest to study
the statistics of the $k$-th maximum $M_{k,N}$ and the statistics of the 
gap defined by $d_{k,N}=M_{k,N}-M_{k+1,N}$. As before, we consider the parent distribution
to be $p(x)$. Then we define the upward and the downward 
cumulative distributions respectively
\bea
p_{>}(x)=\int_{x}^{\infty}p(y)dy \;, \\
p_{<}(x)=\int_{-\infty}^{x}p(y)dy \;.
\eea
Then the cumulative probability $Q_{k,N}(x)$
that the $k$-th maximum stays below $x$ is given by
\bea
Q_{k,N}(x)=\text{Prob}[M_{k,N} \leq x]=\sum_{m=0}^{k-1}{N \choose m}~[p_{>}(x)]^{m}[p_{<}(x)]^{N-m}~,
\eea
where $m$ is the number of points above $x$. The exact PDF is then given by
\bea\label{eq:PkN}
P_{k,N}=\text{Prob}[M_{k,N}=x]=N {N-1 \choose k-1}~p(x)~[p_{<}(x)]^{N-k}[p_{>}(x)]^{k-1} \;.
\eea
{With some more work, one can also obtain the probability distribution of the gap $d_{k,N} = M_{k,N} - M_{k+1,N}$ as (see e.g.~\cite{SM13})}
\bea
P_{k,N}(d)=\text{Prob}[d_{k,N}=d]=N(N-1){N-2 \choose k-1} \int_{-\infty}^{\infty}dx~p(x)p(x-d)[p_{<}(x)]^{N-k-1}[p_{>}(x)]^{k-1} \;.
\eea
As for the cumulative distribution of the maximum, we can again do the same analysis as before and 
extract the leading large $N$ behavior. One finds~\cite{SM13,ABN,ND}
\bea
Q_{k,N}(x)=\text{Prob}[M_{k,N} \leq x]\xrightarrow[z=(x-a_{N})/b_{N}~\text{fixed}]{x \to \infty,~N\to \infty}G_{k}\left(\f{x-a_{N}}{b_{N}}\right)~,
\eea
where the scaling function $G_{k}(z)$ is given by 
\bea\label{Gk}
G_{k}(z)=F_\mu(z) \sum_{j=0}^{k-1}\f{[-\ln F_\mu(z)]^{j}}{j!}= \frac{1}{\Gamma(k)}\,
\int_{-\ln F_\mu(z)}^{\infty} \e^{-t}\, t^{k-1}\, dt~,
\eea
where $F_{\mu}(z)$, with $\mu=1,2,3$,  denote respectively the Fr\'echet, Gumbel
and Weibull scaling functions discussed above respectively in Eqs. (\ref{F1}), (\ref{F2}) and (\ref{F3}). In particular setting $k=1$ (global maximum) in Eq. (\ref{Gk}) one recovers $G_1(z)= F_\mu (z)$, as expected. As an example,
if we choose the parent distribution $p(x)$ from the subclass B where $F_2(z)=\e^{-\e^{-z}}$ 
is the Gumbel distribution, we find
$G_{k}(z)=\e^{-\e^{-z}}\sum_{j=0}^{k-1}\f{\e^{-jz}}{j!}$ and the PDF is $G_{k}^{\prime}(z)=\f{\e^{-kz-\e^{-z}}}{(k-1)!}$.
This is often known as the generalized Gumbel law. Similarly, one can also derive the limiting scaling distributions of the $k$-th gap, $d_{k,N}= M_{k,N}-M_{k+1,N}$ (see Ref.~\cite{SM13}
for more details).

\section{Correlated random variables}
In this section we study the extreme statistics of correlated random variables. 
In the first subsection we revisit 
the random variables where the {correlations are weak} and then we study the ``strongly'' interacting random variables.
There is no general framework to study the statistics of the correlated random variables.
However, we show in the following that for weakly correlated variables, one can provide a rather general renormalization group type of argument to study the extreme statistics. This argument does not work when the variables are strongly
correlated and one has to study case by case different models to
gain an insight.

\subsection{Weakly correlated random variables} \label{Sec:weakly}

\begin{figure}
\includegraphics[width = 0.6\linewidth]{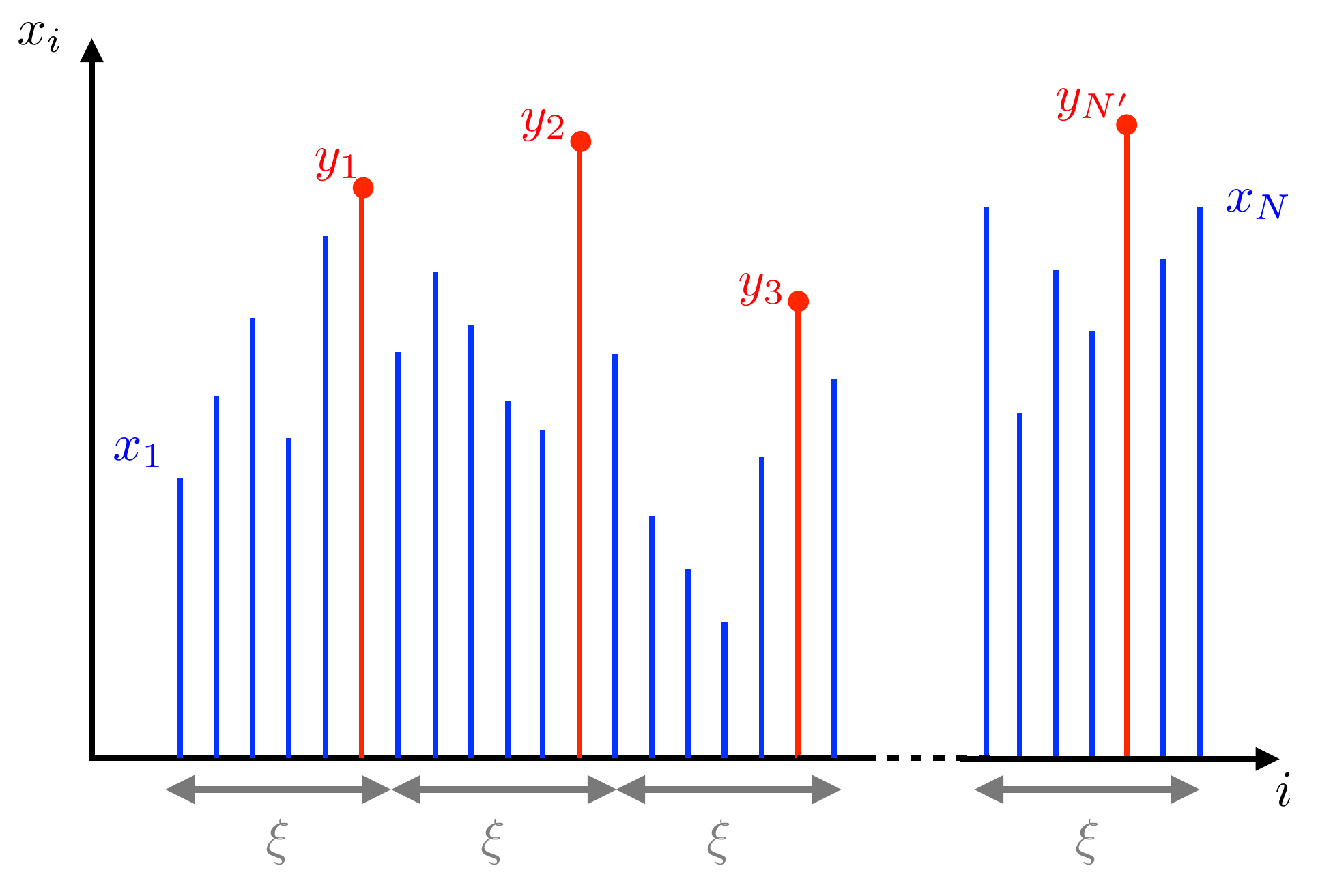}
\caption{Illustration of a set of weakly correlated random variables, characterized by a finite correlation length
$\xi \ll N$, as in Eq. (\ref{correl_weak}). The system is divided in $N' = N/\xi$ blocks of size $\xi$, which are roughly independent. Within each block, we denote by $y_i$'s, with $i=1, \ldots, N'$ the local maxima, which are essentially uncorrelated. Hence, for large $N$, the global maximum $M = \text{Max}[y_{1},y_{2}, \ldots,y_{N'}]$ [see Eq. (\ref{max_weak})] behaves as the maximum of a set of $N'$ IID random variables.}\label{Fig_weak}
\end{figure}

Suppose that we have a set of random variables that are not independent, but correlated
such that the connected part of the correlation function decays fast (say exponentially)
over a certain finite correlation length $\xi$ (see Fig. \ref{Fig_weak})
\bea \label{correl_weak}
C_{i,j}=\langle x_{i}x_{j} \rangle-\langle x_{i}\rangle\langle x_{j}\rangle\sim~\e^{-|i-j|/\xi} \;.
\eea 
Clearly, when two variables are separated over a length scale larger than $\xi$, i.e.,
when $|i-j|\gg\xi$, then they essentially get uncorrelated. 
Now weak correlation implies that $\xi \ll N$, where
$N$ is the total size of the sample.

For such weakly correlated variables one can construct a heuristic argument
to study the extreme statistics~\cite{airy}, as we describe now.
Consider $N'=\xi \ll N$ and break the system into identical blocks
each of size $\xi$ (see Fig. \ref{Fig_weak}).
There
are thus $N/\xi$ number of blocks. 
While the random variables inside each box are still strongly correlated,
the variables that belong to different boxes are approximately uncorrelated.
So, each of these boxes are 
non-interacting. 
Now, for each box $i$, let $y_i$ denote the ``local maximum'', i.e., the maximum
of all the $x$-variables belonging to the $i$-th block, where $i=1,2,\ldots, N'=N/\xi$ (see Fig. \ref{Fig_weak}).
By our approximation, the variables $y_i$'s are thus essentially {\em uncorrelated}.
Hence, we have
\bea\label{max_weak}
M=\text{max}[x_{1},x_{2}, \ldots,x_{N}]=\text{max}[y_{1},y_{2}, \ldots,y_{N'}] \;.
\eea
Therefore, in principle if one knows the PDF of the $y_i$'s, then this problem is essentially
reduced to calculating the maximum of $N'$ uncorrelated random variables
$\{y_1,y_2,\ldots, y_{N'}\}$, which has already been discussed before.
So, we know that depending on the tail of $p(y)$, the limiting distribution
of $M$ of $N$ weakly correlated variables will, for sure, belong to
one of the three (Fr\'echet, Gumbel or the Weibull class)
limiting extreme distributions of IID random variables.
To decide the tail of $p(y)$, of course one needs to solve a {\em strongly}
correlated problem since inside each block the variables are strongly correlated.
However, one can often guess the tail of $p(y)$ without really solving for
the full PDF of $p(y)$ and then one knows, for sure, to which class the
distribution of the maximum belongs to. As a concrete example
of this procedure for weakly correlated variables, we discuss in the next section the Ornstein-Uhlenbeck
stochastic process where one can compute the EVS exactly and demonstrate that
indeed this heuristic renormalization group argument works very well. {Note also that, in the case
of Gaussian random variables $x_i$'s with stationary correlations, i.e. $C_{i,j} = c(|i-j|)$ in Eq. (\ref{correl_weak}), one can show \cite{Ber64} that if $c(x)$ decays faster than $1/\ln(x)$ for large $x$, then the distribution of $x_{\max}$ is still given by a Gumbel distribution.}

To summarize, the problem of EVS of weakly correlated random variables 
basically reduces to IID variables with an effective number $N'=N/\xi$ where
$\xi$ is the correlation length. So, the real challenge is to compute
the EVS of strongly correlated variables where $\xi\geq O(N)$, to which
we now turn to below.

\subsection{Strongly correlated random variables}

Strongly correlated means that the correlation length prevails over the whole system such that the idea of block 
spins as in Fig. \ref{Fig_weak} will no longer hold. A general theory for calculating the
EVS, such as in case
of IID or weakly correlated variables, is currently lacking
for such strongly correlated variables. In the absence of a general theory,
one tries to study different exactly solvable special cases in the hope
of gaining some general insights. There are examples, though their numbers
are unfortunately few, where the EVS of a strongly correlated system can be computed
exactly. In the following, we will discuss a few examples of such
exactly solvable cases.

As a first example of a strongly correlated system, we will consider the one-dimensional
Brownian motion, for which the distribution of maximum can be computed explicitly.
Next, we will discuss the Ornstein-Uhlenbeck (OU) process, which
represents the noisy motion of a classical particle in a harmonic well. We will see
that the OU process actually represents a weakly correlated system, for which one can
compute the distribution of the maximum explicitly, demonstrating the power
of the heuristic argument presented in the previous subsection for weakly
correlated variables. Next, we will discuss the case of EVS of random variables subjected to
a global constraint, like the famous {\it zero-range} process, and then the case of hierarchically and logarithmically
correlated variables. We will then turn to the EVS
in stochastic resetting systems, after which we will 
present the problem of the maximum height distribution of a fluctuating $(1+1)$-dimensional 
interface in its stationary state.
Finally, we will discuss the statistics of the largest eigenvalue for Gaussian random matrices.

\subsubsection{One-dimensional Brownian motion and random walks}

\begin{figure}[t]
\includegraphics[width=0.6\linewidth]{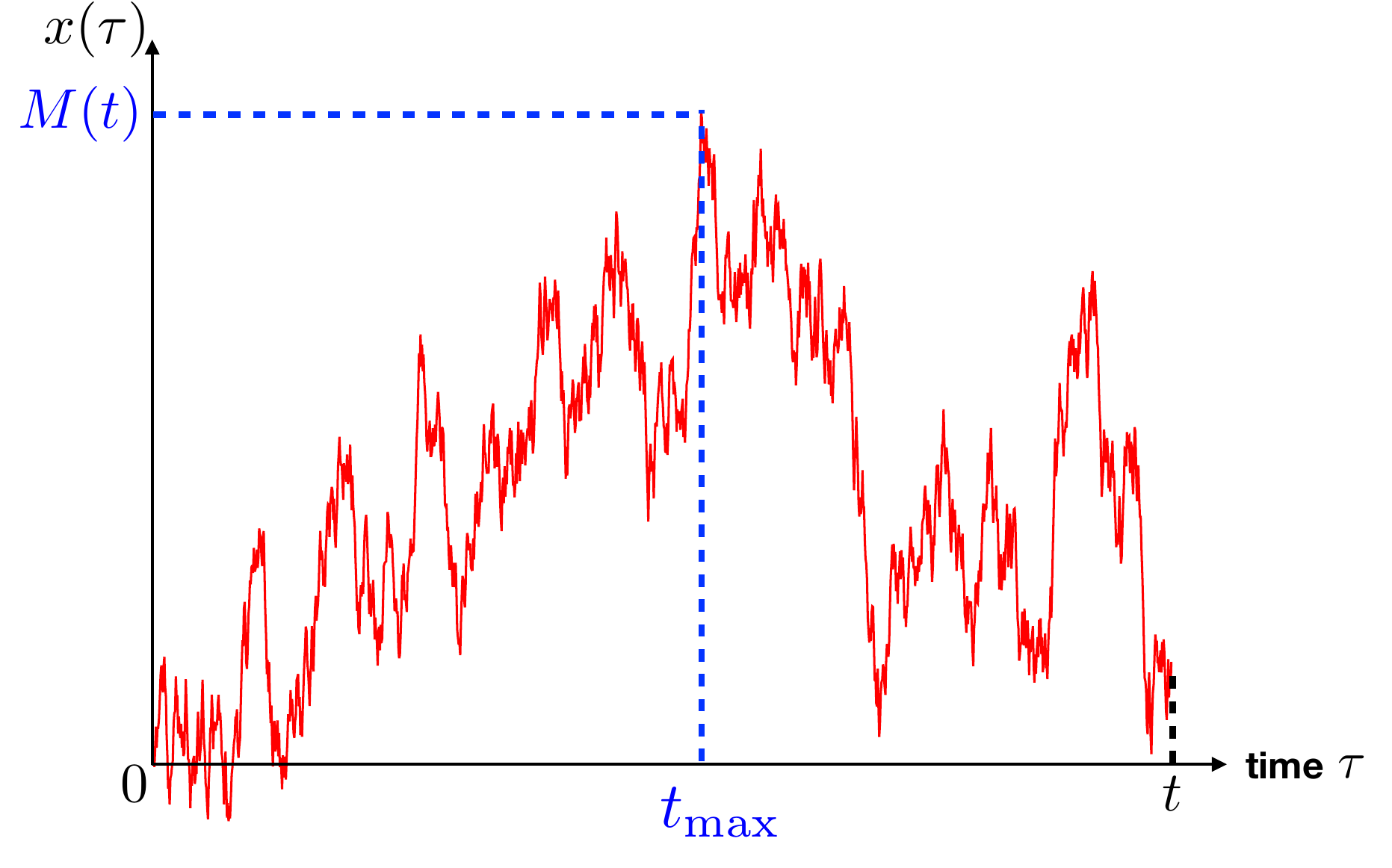}
\caption{Trajectory of a Brownian motion $x(\tau)$ as in Eq. (\ref{def_BM}) over the time interval $[0,t]$. The main extreme observables are the maximum $M(t)$ as well as the time $t_{\max}$ at which the maximum is reached.}\label{Fig:BM}
\end{figure}

We consider the case where the random variables $x_i$'s represent the positions
of a one-dimensional random walker at discrete time-step $i$, starting from $x_0=0$.
We are interested in the maximum position of the walker up to step $n$. Even though
the discrete problem can be solved explicitly (see the discussion below), for simplicity we will first consider
below the continuous-time version of the random walk, i.e., a one-dimensional
Brownian motion whose position $x(\tau)$ evolves via the stochastic Langevin equation
\bea \label{def_BM}
\f{dx}{d\tau}=\eta(\tau) \;,
\eea
starting from $x(0)=0$ and
$\eta(\tau)$ represents a Gaussian white noise with zero mean $\langle \eta(\tau) \rangle=0$ 
and delta-correlations $\langle \eta(\tau)\eta(\tau') \rangle=2D~\delta(\tau-\tau')$. 
We are interested in the
PDF of the maximum $M(t)$ of this Brownian motion $x(\tau)$ over the time window $[0,t]$ (see Fig. \ref{Fig:BM})
\bea
M(t) = \max_{0\le \tau\le t}\left[x(\tau)\right] \;. \label{def_max}
\eea
To proceed, we first note that $x(\tau)= \int_0^{\tau} \eta(s)\, ds$ and hence, one can easily
compute the mean and the correlator of the process $x(\tau)$: 
$\langle x(\tau) \rangle=0$ and $\langle x(\tau)x(\tau') \rangle=2D~\text{min}(\tau,\tau')$.
Thus the variables $x(\tau)$ at different times are strongly correlated. The
correlation function does not decay and persists over the whole sample size $\tau\in [0,t]$.

{As done in the IID case [see Eq. (\ref{def_cumul})}] it is also very useful here to define the 
cumulative distribution of the maximum  
\bea
Q(z,t)&=&\text{Prob}[M(t)\leq z] = \text{Prob}[x(\tau)\leq z,~0 \leq \tau \leq t] \;.
\eea
{First, we notice that $Q(z,t) = 0$ for $z<0$ since $M(t) \geq x(0) = 0$.} To compute $Q(z,t)$ for $z \geq 0$, we note that it just represents the
probability that the Brownian particle, starting at $x(0)=0$, stays below
the level $z$ up to time $t$. Let $P(x,t|z)$ denote the
probability density for the particle to reach $x$ at time $t$, while staying below
the level $z$ during $[0,t]$. It is then easy to see that $P(x,t|z)$ satisfies
the diffusion equation in the semi-infinite domain $x\in (-\infty, z]$ 
\bea \label{diff_eq}
\f{\partial P}{\partial t}&=&D\f{\partial^{2}P}{\partial x^{2}}~,
\eea 
with the following initial and boundary conditions
\bea \label{bc}
P(x,0|z)&=&\delta(x)~, \nn \\
\text{and~}P(x\to -\infty,t|z)&=&0,~P(x=z,t|z)=0\,.
\eea
The absorbing boundary condition $P(x=z,t|z)=0$ at $x=z$ guarantees that
the particle does not cross the level at $x=z \geq 0$. 
{The solution is simple and can be obtained by the method of images~\cite{redner,satya_review10,BMS13}
\bea
P(x,t|z)=\f{1}{\sqrt{4\pi Dt}}[\e^{-\f{x^{2}}{4Dt}}-\e^{-\f{(x-2z)^{2}}{4Dt}}] \;.
\eea
Therefore the cumulative distribution function of the maximum is given by (for $z \geq 0$)
\bea \label{CDF}
Q(z,t)=\int_{-\infty}^{z}dx~P(x,t|z)={\rm erf}\left(\f{z}{\sqrt{4Dt}}\right);\,\, {\rm where}\,\, 
{\rm erf}(z)= \frac{2}{\sqrt{\pi}}\,\int_0^z \e^{-u^2}\,du \;, \\
\eea
while we recall that $Q(z,t) = 0$ for $z<0$. Hence the PDF $P_M(z,t)$ of $M(t)$ is given by
\bea \label{PDF_M}
P_M(z,t)=\frac{\partial Q(z,t)}{\partial z}=\theta(z)\f{1}{\sqrt{\pi Dt}}\e^{-\f{z^{2}}{4Dt}} \;,
\eea
where $\theta(z)$ is the Heaviside theta function, i.e. $\theta(z) = 1$ if $z>0$ and $\theta(z) = 0$ if $z<0$. In particular, from Eq. (\ref{PDF_M}), one easily computes the expected maximum $\langle M(t) \rangle = \frac{2}{\sqrt{\pi}}\,\sqrt{D\,t}$.
This thus represents, perhaps, the simplest example of a strongly correlated system
for which one can compute the distribution of the maximum exactly.} 

{In fact various other extreme observables can 
be computed for Brownian motion and its variants (like the Brownian bridge, which is the BM conditioned to start and end at the origin at time $t$, or the Brownian excursion, which is a Brownian bridge conditioned, additionally, to stay positive on $[0,t]$). This includes for instance the temporal correlations of the (running) maximum \cite{BKMO16a,BKMO16}, the time $t_{\max}$ (see Fig. \ref{Fig:BM}) at which the BM or its variants reaches its maximum $M(t)$ \cite{MRKY08,SL10,Lev40} (see section \ref{sec:tM} below for a more complete discussion of this observable) as well as other functionals of $M(t)$~\cite{PCMS13, PCMS15}. These extremal properties of BM have found several interesting applications like in the statistics of the convex hull of {\it two-dimensional} BM \cite{RMC09,MCR10,CBM15}, in enumerative combinatorics and computer science \cite{BKR72,Maj05,PCMS13,PCMS15,CMY03} as well as in the study of fluctuating interfaces \cite{Satya:04,airy}, which we will discuss further below. EVS have also been studied for more general stochastic processes (i.e. beyond Brownian motion and its variants) involving (i) either a single degree of freedom, like the so called continuous time random walks (CTRW) \cite{SL10,FM12}, the random acceleration process \cite{BGMZ07,MRZ10} and more general $1/f^\alpha$ noise \cite{GMOR07,MOR11,RBKS11} as well as  fractional Brownian motion \cite{Mol99,DW16,DW16b,DRW17,SDW18} (ii) or several degrees of freedom like $N$ independent \cite{KMR10,KMS13} as well as non-intersecting Brownian motions \cite{KIK08,SMCR08,KIK08b,Fei09, RS11,FMS11,Lie12,SMCF13,KMS14, NR17,Tristan} and branching Brownian motions~\cite{SF79, BD1,BD2,DMRZ13,RMS14,RMS15,RMS15b,DMS16,DS16,DS17}. Extreme value questions for these processes have recently found 
many applications ranging from the number of common and distinct visited sites by $N$ random walkers \cite{KMS13}, distribution of the 
cover time by $N$ Brownian motions on an interval of length $L$ \cite{MMS16} and the distribution of the cover time of a single transient walker in higher dimensions \cite{CBV15}, combinatorics \cite{KIK08,SMCR08,Fei09}, fluctuating interfaces in the Kardar-Parisi-Zhang universality class \cite{SMCR08,RS11,FMS11,SMCF13,BLS12}, disordered systems \cite{BD1, BD2}, genetics \cite{SF79}, propagation of epidemics \cite{DMRZ13} or in random planar geometry \cite{HCS08, RMR11,LGE13}.}

{All these properties discussed so far in this section concern continuous time processes, but one can also study extreme value questions for discrete time processes, like random walks on the real line mentioned in the introduction of this section. In this case, at each time step, the walker performs a jump $\eta$ whose size is a random variable drawn from a jump PDF $f(\eta)$. The position of the walker $x_n$ thus evolves according to the Markov dynamics
\bea \label{eq:RW}
x_n = x_{n-1} + \eta_n \;,
\eea
starting from a given initial position $x_0$. In Eq. (\ref{eq:RW}) the $\eta_n$'s are IID random variables distributed according to $f(\eta)$. The RW model (\ref{eq:RW}) includes discrete random walks (RWs), e.g. $f(\eta) = (1/2) \delta(\eta - 1) + (1/2) \delta(\eta+1)$, as well as RWs with continuous jump distributions, such as a uniform, a Gaussian or a Cauchy distribution for instance. One usually distinguishes between two different types of RWs: (i) the ``standard'' RWs, for which $\sigma^2 = \langle \eta^2 \rangle - \langle \eta \rangle^2$ is finite, which converge after a large number of steps to the Brownian motion as in Eq. (\ref{def_BM}) and (ii) the L\'evy flights, for which $\sigma$ is infinite, which is the case for heavy tailed distributions $f(\eta) \propto |\eta|^{-1-\mu}$ for large $|\eta|$, with a so-called L\'evy index $0<\mu<2$ (for instance, a Cauchy jump distribution corresponds to $\mu=1$). For large $n$, the extreme value properties of the standard RWs coincide at leading order with that of the BM studied above while they are quite different for the L\'evy flights. Unlike the Brownian motion which is continuous both in space and time, the study of EVS for RWs and L\'evy flights is usually much harder. Nevertheless, there is now an important literature on this subject and a lot of results are now available \cite{Feller:71,Spi56,Pol52,SM13,CM05,satya_review10,BMS13,MMS17,MMS18}, which are useful for instance for discrete height model \cite{SM06}, in computer science (e.g., for packing problems in two-dimensions \cite{CFFH98}) or to characterize the convex hull of two-dimensional random walks \cite{GLM2017}. Interestingly, for discrete-time RWs and L\'evy flights described by Eq. (\ref{eq:RW}), one can study the statistics not only of the global (of first) maximum $M(n)$ but also of the second, third, ..., more generally of the $k$-th maximum, as done previously in the IID case [see Eq. (\ref{def_order})]. Of particular interest in this context are the gaps between two successive maxima and for which a lot of results have been obtained during the last years~\cite{Cha99,MOR11,SM12,MMS13,MMS14,MMS16,MS17,BMS17,LMS19}.}


\subsubsection{Ornstein-Uhlenbeck (OU) process}
We now consider a Brownian particle in a harmonic potential governed by the 
following Langevin equation
\bea
\f{dx}{d\tau}=-\mu x+\eta(\tau)~,
\label{ou1}
\eea 
where, as before, $\eta(\tau)$ is a Gaussian white noise with zero mean and
is delta correlated, $\langle \eta(\tau)\eta(\tau')\rangle = 2\,D\, \delta(\tau-\tau')$.
Assuming that the particle starts at the origin $x(0)=0$, $x(\tau)$ is a Gaussian random variable of zero mean at all time $\tau \geq 0$ (since its equation of
evolution (\ref{ou1}) is linear). It is thus fully characterized by its two-time correlation function $C(t_1,t_2)$ which can be easily computed as
\bea
C(t_{1},t_{2})=\langle x(t_{1})x(t_{2}) \rangle=\f{D}{\mu}[\e^{-\mu|t_{1}-t_{2}|}-\e^{-\mu(t_{1}+t_{2})}] \;.
\eea
{Clearly the correlation function reduces to the Brownian limit when $\mu \to 0$, since in this limit 
$C(t_1,t_2) \to 2\, D\, {\rm min}(t_1,t_2)$ and the system becomes strongly correlated.
In contrast, for nonzero $\mu>0$, the correlation function,
at large times $t_{1},t_{2}\gg1/\mu$, decays exponentially with the time-difference
$C(t_{1},t_{2})\approx\f{1}{2\mu}\e^{-\mu|t_{1}-t_{2}|}$
with a correlation length $\xi=1/\mu$. From our above arguments about weakly
correlated random variables (see Section \ref{Sec:weakly}), we would then expect to get the limiting Gumbel
distribution for $\mu>0$ (since $x(\tau)$ is a Gaussian random variable of variance $\sqrt{D/\mu}$ for large~$\tau$). We demonstrate below how this Gumbel distribution emerges by solving exactly the EVS for the OU process.} 

As before, let $Q(z,t)$ denote the cumulative distribution of the maximum
$M(t)$ of the OU process in the time interval $[0,t]$. The particle
starts at the origin $x(0)=0$ and evolves via Eq. (\ref{ou1}).
Let $P(x,t|z)$ denote the probability density for the particle to
arrive at $x$ at time $t$, while staying below the level $z$.
This restricted propagator satisfies the Fokker-Planck equation
in the domain $x\in (-\infty,z]$
\bea
\f{\partial P}{\partial t}&=&D\f{\partial^{2}P}{\partial x^{2}}+\mu \f{\partial}{\partial x}\left( x\,P \right) \;,
\label{fp.ou}
\eea
with the initial condition $P(x,0|z)= \delta(x)$ and the boundary conditions
$P(x,t|z)=0$ as $x\to -\infty$ together with the absorbing condition at level $z$, i.e. $P(x=z,t|z)=0$
for all $t$. For simplicity, we will set $D=1/2$.
We note that, unlike in the Brownian case ($\mu=0$), for $\mu>0$ we can
no longer use the method of images due to the presence of the potential. 
However, one can solve this equation by the eigenfunction expansion and the solution
can be expressed as 
\bea
P(x,t|z)=\sum_{\lambda} a_{\lambda}\, \e^{-\lambda t}\, D_{\lambda/\mu}(-\sqrt{2\mu}x)\, \e^{-\mu\, x^2/2}
\label{sol1_ou}
\eea
where  $D_{p}(z)$ is the 
parabolic cylinder function which satisfies the second order ordinary differential
equation: $D_p''(z)+ (p+1/2-z^2/4)\,D_p(z)=0$ (out of the two linearly independent
solutions, we choose the one that vanishes as $z\to \infty$). The absorbing boundary condition
$P(x=z,t|z)=0$ induces the boundary condition on the eigenfunction
\bea \label{condition_ev}
D_{\lambda/\mu}\left(-\sqrt{2\mu}z\right)=0 \;,
\eea
which then fixes the eigenvalues $\lambda$'s, which are necessarily positive. Finally, in Eq. (\ref{sol1_ou}) the $a_\lambda$'s can be computed explicitly
\bea
a_\lambda = \frac{\sqrt{2 \pi \mu} \, 2^{\frac{\lambda}{2\mu}}}{\Gamma\left( \frac{1-\lambda/\mu}{2}\right) \, c_\lambda} \;, \; c_\lambda = \int_{-\infty}^\infty D^2_{\lambda/\mu}(x) \, \e^{-x^2/2} dx\;.
\eea

At large times $t$, the summation in Eq. (\ref{sol1_ou}) is dominated
by the term involving the smallest eigenvalue ${\lambda}_0 (z)$, which evidently
depends on $z$. For arbitrary $z$, it is difficult to solve 
$D_{\lambda/\mu}(-\sqrt{2\mu}z)=0$ and determine the smallest eigenvalue
$\lambda_0(z)$. However, for large $z$, one can make progress by perturbation
theory and one can show that to leading order for large $z$,
\bea
\lambda_0(z)\xrightarrow{z\to \infty}\f{1}{\sqrt{\pi}}\,\mu^{3/2}\,z\,\e^{-\mu z^{2}} \;,
\label{l0}
\eea
while $a_{\lambda_0} \to a_0 = \sqrt{2\mu/\pi}$ (note also that $D_0(x) = \e^{-x^2/4}$). Consequently, for large $t$ and large $z$,
\bea
Q(z,t)\sim \e^{-\lambda_0(z)\,t}&\sim& \e^{-\e^{-\mu z^{2}+\log (\f{ t \mu^{3/2}z}{\sqrt{\pi}})}} \nn \\
&\rightarrow&F_{2}\left[\sqrt{4\,\mu\, \ln t}\left(z-\frac{1}{\sqrt{\mu}}\, \sqrt{\ln t}\right)\right] \;,
\label{ou_gumbel}
\eea
where $F_{2}(y)=\exp[-\exp[-y]]$ is the Gumbel distribution. 
As a result, for $\mu>0$, the average value of the maximum grows very slowly for large $t$ as,
$\langle M(t)\rangle \sim \frac{1}{\sqrt{\mu}}\, \sqrt{\ln t}$, while its 
width around the mean decreases as $\sim 1/\sqrt{\ln t}$. In fact, for $\mu>0$, a full analysis of the mean value of the maximum $\langle M(t)\rangle$ for all $t$
shows that initially it grows as $\sqrt{t}$ (for $t \ll 1/\mu$) where it does not feel the
confining potential and hence behaves as a Brownian motion. But for $t \gg 1/\mu$, the particle
feels the potential and the mean maximum crosses over to a slower growth as $\sqrt{\ln t}$. Hence, one has
\bea
\langle M(t) \rangle \sim \left\{
     \begin{array} {rl}
     {\sqrt{t}}     & ~ \text{for~} t \ll1/\mu \\
     {\sqrt{\ln t}} & ~ \text{for~} t \gg 1/\mu \;,\\
     \end{array} \right.
\eea
which indicates a crossover from a strongly correlated regime at short time $t \ll 1/\mu$ to an IID regime at longer time $t \gg 1/\mu$.

\subsubsection{Extreme statistics in the presence of a global constraint}

Another interesting example of a correlated system is when the correlations emerge due to the presence of
a global conservation law for the dynamics of the underlying random variables. There are several examples
of this that we briefly discuss here. 

The first example concerns a well known interacting particle systems known as the zero range process (ZRP) \cite{EH2005,Satya_review_ZRP} where one considers a one-dimensional lattice of $N$ sites with periodic
boundary condition. At each site, there is a certain number of particles $\ell_i \geq 0$. From the $i$-th site, a single
particle hops to a neighbouring site with a rate $U(\ell_i)$ that depends only on the occupation number of the 
departure site, provided $\ell_i > 0$. Clearly, the dynamics conserves the total number of particles $\sum_{i=1}^L \ell_i = L$. 
At long times, the system reaches a stationary state where the joint distribution of the occupation numbers at different sites
is given by
\begin{eqnarray}\label{eq:ZRP}
P_{\rm ZRP}(\ell_1,\cdots,\ell_N|L) = \frac{1}{Z_N(L)} \prod_{i=1}^N f(\ell_i) \, \delta_{\sum_{i=1}^N \ell_i,L} \;,
\end{eqnarray} 
where $f(\ell) = 1/\prod_{k=1}^\ell U(k)$ and $Z_N(L)$ is the normalization constant. The $\delta$-function in (\ref{eq:ZRP}) enforces the
global constraint that the number of particles $L$ is fixed. Without this $\delta$-function, the joint distribution would factorize
and we would be back to the IID case. However, the presence of this $\delta$-function makes these variables correlated.
We are interested in the extreme statistics, i.e. the statistics of $\ell_{\max} = \max_{1 \leq i \leq N} \ell_i$. This corresponds
to the largest number of particles carried by a single site, out of $N$ sites. Of particular physical interest is the case when
the transfer rate function $U(k)$ is such that the steady state weight function has a tail such that $\e^{-c\ell} \ll f(\ell) \ll 1/l^2$ for large $\ell$, where $c$ is any arbitrary constant \cite{Satya_review_ZRP}. A well studied example is the case when $f(\ell)$ has a power law tail, $f(\ell) \sim A/\ell^\gamma$ for large $\ell$, with $\gamma >2$ and $A>0$. In this case, it is known that a condensation transition happens when the density $\rho = L/N$ exceeds a certain critical value
$\rho_c$. For $\rho < \rho_c$ the particles are ``democratically'' distributed among all sites: this is the ``fluid'' phase. 
For $\rho > \rho_c$ a single
condensate site emerges that carries a finite fraction of the total number of particles: this is the ``condensed'' phase. Clearly, in this condensed phase, $\ell_{\max}$ is identified
with the number of particles carried by this single condensate.

This ZRP is one example where, despite the presence of correlations, the
statistics of $\ell_{\max}$ can be computed explicitly, in the limit of large $N$, large $L$, but with the density $\rho=L/N$ fixed \cite{Satya:08}. Let us summarize the main results. For any fixed $\gamma > 2$, one finds that the center and scaled distribution of $\ell_{\max}$, in the fluid phase $\rho<\rho_c$, converges to the Gumbel distribution discussed before in Section \ref{sec:IID}. In this case, the global constraint enforced by the $\delta$-function can be relaxed (akin to canonical to grand-canonical) and one can replace it in Eq. (\ref{eq:ZRP}) by $\e^{-\mu \sum_{i=1}^N \ell_i}$ where $\mu > 0$ is the chemical potential (or Lagrange multiplier). This factorizes the joint distribution in Eq. (\ref{eq:ZRP}) into a product of effective IID random variables with an effective distribution $f_{\rm IID}(\ell) = f(\ell)\, \e^{-\mu \ell}$. Thus, for $\mu >0$, the distribution of these IID variables has an effective exponential tail and, hence, the distribution of their maximum converges to the Gumbel distribution, as discussed in Section \ref{sec:IID}. Note that this exponential tail is induced here by the global constraint. When $\rho$ approaches $\rho_c$ from below, $\mu \to 0$ and we again obtain a factorization of the joint distribution in Eq. (\ref{eq:ZRP}) but now the effective distribution has a power law tail $f(\ell) \sim A/\ell^\gamma$ for large $\ell$. Consequently, the distribution of the maximum, centred and scaled, approaches the Fr\'echet distribution, as discussed in Section~\ref{sec:IID}, thus as if the global constraint does not play any role at all at the critical point $\rho = \rho_c$. However, for $\rho > \rho_c$, it turns out that the global constraint has a much stronger effect and the scaled distribution of $\ell_{\max}$ has a non-trivial distribution given by~\cite{Satya:08}
\begin{eqnarray}\label{PDF_lmax1}
{\rm Prob.}(\ell_{\max} = x,N) \sim \frac{1}{N^{\delta}} V_\gamma \left( \frac{x-(\rho-\rho_c)N}{N^\delta} \right)~,
\end{eqnarray}    
where the exponent $\delta$ is given by
\begin{eqnarray}\label{exp_delta}
\delta = 
\begin{cases} 
&\frac{1}{\gamma-1} \;, \quad\quad 2 < \gamma < 3 \;, \\
&\\
& 1/2 \;, \quad \quad \quad \quad \gamma > 3 \;.
\end{cases}
\end{eqnarray}
The scaling function $V_\gamma(z)$ also depends on $\gamma$. While for $\gamma>3$ it is a simple Gaussian, for $2 < \gamma < 3$, it has a non-trivial form given by the integral representation
\begin{eqnarray}\label{V_Laplace}
V_\gamma(z) = \int_{- i \infty}^{+ i \infty } \frac{ds}{2\pi i} \e^{-z s + b s^{\gamma-1}} \;,
\end{eqnarray}
where $b = A \Gamma(1-\gamma)$ with $A$ being the amplitude of the power law tail of $f(\ell) \sim A/\ell^{\gamma}$ for large $\ell$. This therefore represents an example where correlations affect the extreme value distribution non-trivially and yet remains computable.

Another example with a global constraint concerns the largest time interval between
returns to the origin of a one-dimensional lattice random walk \cite{FIK95} and more generally a renewal process \cite{GMS2009,GMS2015}. In this case, let $\ell_i$ denote the time interval between the $i$-th and the $(i-1)$-th return to the origin of a random walk of $L$ steps. After every return to the origin, the process gets renewed due to the Markov nature of the walk. Consequently, 
the joint PDF of the intervals between  successive returns and the total number of intervals $N$ in $L$ steps is given by
\begin{eqnarray}\label{eq:renewal}
P_{\rm REN}(\ell_1,\cdots,\ell_N,N|L)  \propto \left[\prod_{n=1}^{N-1} f(\ell_n)\right] q(\ell_N) \, \delta_{\sum_{n=1}^N \ell_n,L} \;,
\end{eqnarray}
where $L$ represents the total time interval. Here the number of intervals $N$ is also a random variable, unlike in the ZRP case (\ref{eq:ZRP}) where $N$ was fixed. In addition, at variance with Eq. (\ref{eq:ZRP}), while the first $N-1$ intervals in Eq. (\ref{eq:renewal}) 
have the same weight $f(\ell_n)$, the last one has a different weight $q(\ell_N) = \sum_{\ell=\ell_N+1}^\infty f(\ell)$ (this is due the fact that the last interval is yet to be renewed).  In these renewal processes (\ref{eq:renewal}), the weight $f(\ell)$ is taken as an input in the model. For example, for a $1d$-random walk, $f(\ell) \sim 1/\ell^{3/2}$ for large $\ell$. The statistics of $\ell_{\max} = \max_{1 \leq i \leq N} \ell_i$ has been studied in detail in Refs. \cite{GMS2009,GMS2015}. The main results can be summarized as follows. If $f(\ell)$ decays faster than $1/\ell^2$ for large $\ell$, one essentially recovers the IID results for the distribution of $\ell_{\max}$, indicating that the global constraint does not play any significant role. However, for $f(\ell) \sim 1/\ell^{\gamma}$ with $1 < \gamma < 2$, the constraint significantly modifies the distribution of $\ell_{\max}$, leading to a nontrivial distribution \cite{GMS2015}. In addition, in Refs. \cite{GMS2009, GMS2015}, another extreme observable was studied, namely the probability that the last interval is the longest.

We finish this section with yet another example of extreme statistics in the presence of a global constraint. It appeared in a recently introduced
truncated long-ranged Ising model in a one-dimensional lattice of size $L$,  where the couplings between two spins at sites $i$ and $j$ decay as an inverse square law, provided both spins belong to the same domain. The spins across domains do not have any long range interaction \cite{BM2014,BM2014b}. In addition to this long-range interaction, there is also a short-range interaction between neighboring spins. This model was shown to exhibit a mixed order phase transition (MOT) at a finite critical temperature where the correlation length diverges, as in a second order phase transition, but the order parameter undergoes a jump as in a first order phase transition. The configurations of the system can be labelled by the lengths of spin domains $\{\ell_i \}_{1 \leq i \leq N}$ where $N$ represents the number of domains. 
At finite temperature the Boltzmann weight associated to such a configuration can be written as
\begin{eqnarray}
P_{\rm MOT}(\ell_1, \ell_2, \cdots, \ell_N,N|L) = \frac{1}{Z(L)}\, y^N \, \prod_{n=1}^N \frac{1}{\ell_n^c} \, {\delta}_{\sum_{n=1}^N \ell_n, L} \,, \label{eq:jpdf}
\end{eqnarray}
where $\delta_{i,j}$ is the usual Kronecker delta and $c>1$ is related to both the temperature and the long-range coupling constant. The constant $y$ represents the fugacity which controls the number of domains $N$, which actually fluctuates from one configuration to another. Once again, there is a global constraint enforcing the sum rule that the domain lengths add up to the system size $L$.  Again, this model differs slightly from the two models discussed above, namely the ZRP (\ref{eq:ZRP}) and the renewal process (\ref{eq:renewal}). In the $(y,c)$ plane, this model has a phase transition across the critical line $y_c = 1/\zeta(c)$ where $\zeta(x)$ is the Riemann zeta function. For $y>y_c$ the system is in a paramagnetic phase with a large number of domains. In contrast, for $y<y_c$, the system is ferromagnetic with one large domain of size $\propto L$. Very interesting properties emerge on the critical line $y=y_c$. In this case, for $1<c<2$, the largest domain size $\ell_{\max}$ scales extensively with the system size, $\ell_{\max} \propto L$ and the PDF of $\ell_{\max}$ exhibits a scaling form \cite{BMSM2016}
\begin{eqnarray}\label{scaling_Fc}
{\rm Prob.}(\ell_{\max} = x,L) \approx \frac{1}{L} F_c\left( \frac{x}{L}\right) \;,
\end{eqnarray} 
where the scaling function $F_c(z)$ was computed exactly and was found to exhibit non-analytic behaviors at $z=1/k$ where $k$ is a positive integer. This scaling form (\ref{scaling_Fc}) also indicates that the fluctuations are anomalously large in this regime. This fact was shown to be related to the so-called fluctuation-dominated phase ordering (FDPO) regime where the spin-spin correlation function develops a cusp at short distance \cite{BMM2019}. Thus we see that, in this regime $1<c<2$ on the critical line $y=y_c$, the global constraint in Eq. (\ref{eq:jpdf}) does affect the distribution of $\ell_{\max}$ in a nontrivial way. In contrast, for $y=y_c$ but $c>2$, one recovers the Fr\'echet distribution of IID random variables, indicating that the global constraint is insignificant to a large extent (it only renormalises the scale factors \cite{BMSM2016}).

  \subsubsection{Hierarchically and logarithmically correlated random variables}

\begin{figure}[t]
\includegraphics[width = 0.3\linewidth]{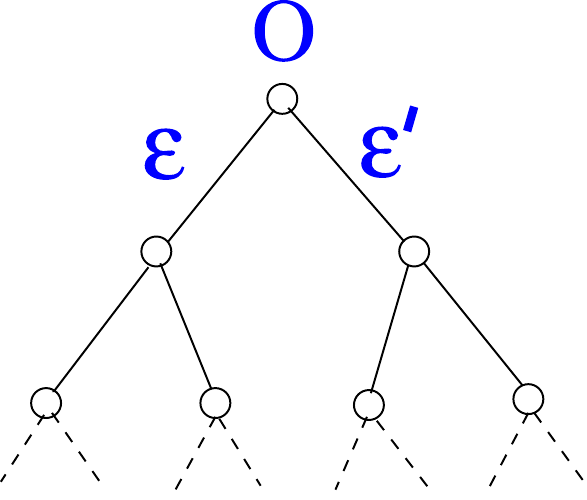}
\caption{A binary tree with a root at $O$. On each bond $i$, an energy $\epsilon_i$ is assigned, each drawn independently from a distribution $\rho(\epsilon)$. For example, in the figure, we show the energies $\epsilon$ and $\epsilon'$ of the two daughter bonds of $O$. A directed polymer starts at $O$ and goes down the tree to the leaves and the energy of the path is given by the sum of the disordered bond energies belonging to the path.}\label{Fig_dp}
\end{figure}

So far, we have been discussing the extreme statistics of uncorrelated, weakly
correlated and strongly correlated random variables. There exists however a class of problems where the variables are
{\it hierarchically} correlated. This is typically the case for disordered models defined on a tree. For example, one can consider the celebrated
problem of a directed polymer on a tree \cite{DS1988}. Consider a rooted Cayley tree where each node has two off-springs. On each
bond of the tree with $n$ generations, starting from a single root at $O$, we assign a positive quenched random variable $\epsilon_i \geq 0$ (see Fig. \ref{Fig_dp}). A configuration of $\{ \epsilon_i\}$'s specifies the disordered environment. We then consider a directed polymer of $n$ steps that goes from the root to one of the leaf sites at the $n$-th generation. The energy of such a directed path is given by the sum of the energies of all the bonds belonging to the path
\begin{eqnarray} \label{energy_dp}
E_{\rm path} = \sum_{i \, \in {\rm \, path}} \epsilon_i    \;.
\end{eqnarray} 
Thus there are $N = 2^n$ possible paths, each with its own energy. This model was initially studied at finite temperature $T>0$, where it was found that 
it exhibits a transition to a spin-glass like (``frozen'') phase at low temperature (reminiscent of a replica symmetry breaking scenario found in mean-field spin-glasses) \cite{DS1988}. 
This problem was then revisited at $T=0$ \cite{Satya:00, Dean:01} where  
one is interested in the path that has the 
{\it minimal} energy
\begin{eqnarray} \label{min_dp}
E_{\rm min} = \min \left[ {E_{\rm path}}_1, {E_{\rm path}}_2, \cdots, {E_{\rm path}}_{N} \right] \;.
\end{eqnarray} 
This is then a typical extreme value problem and clearly, 
because of the partial overlaps between the paths, the energy variables $\{{E_{\rm path}}_i \}$'s are correlated. Indeed the correlations
between the energies of any two paths is proportional to the number of bonds they have in common: for these reasons, the $\{{E_{\rm path}}_i \}$'s are called {\it hierarchically correlated random variables}. It was shown in a series of work that the EVS of such random variables is closely related to traveling fronts \cite{Satya:00,Dean:01,KM00,MK03,Satya:02,Satya_Ziv,BenNaim}. Indeed, in these cases, the cumulative distribution of the minimum (or the maximum) is generically a traveling front solution of a non-linear equation, the most well--known example being the celebrated Fisher - Kolmogorov, Petrovsky, Piskunov (F-KPP) equation. Within this connection, the number of generations $n$ plays the role of time and 
the average value $\langle E_{\rm min} \rangle$ corresponds to the position of the front. By using the tools developed to study such traveling front solutions, in particular the ``velocity selection principle'' that provides a criterion to select the velocity of the traveling front, one can easily compute the two leading terms (in the limit of large $n$) of the average value of the minimum (or the maximum). 

For the directed polymer on the Cayley tree with random energies $\epsilon_i$'s distributed according to a PDF $\rho(\epsilon)$, one can derive 
a recursion relation for the CDF of $E_{\min}$ in (\ref{min_dp}), $P_n(x) = {\rm Prob.}(E_{\min} \geq x)$, which reads \cite{Satya:00}
\begin{eqnarray}\label{rec_DPRM}
P_{n+1}(x) = \left[\int \rho(\epsilon) P_n(x-\epsilon) \, d\epsilon \right] \, \left[\int \rho(\epsilon') P_n(x-\epsilon') \, d\epsilon' \right] =\left[\int \rho(\epsilon) P_n(x-\epsilon) \, d\epsilon \right]^2 \;,
\end{eqnarray}
together with the initial condition $P_0(x) = \theta(-x)$. This relation follows from the following reasoning. Consider a particular realization of the disordered bond energies $\epsilon_i$'s on a tree, as in Fig. \ref{Fig_dp}. The probability $P_n(x) = {\rm Prob.}(E_{\min} \geq x)$ is also the probability that all paths originating at $O$ have energies bigger than $x$. Any path originating from the root $O$ either passes through the left bond with energy $\epsilon$ or through the right bond with energy $\epsilon'$ as in Fig. \ref{Fig_dp}. The left (respectively right) subtree starting with the left (respectively right) daughter of $O$ must have energy bigger that $x-\epsilon$ (respectively $x-\epsilon'$). Hence, the probability these joint events, using the fact that the sub-trees are statistically independent, is given by the product $P_n(x-\epsilon) P_n(x-\epsilon')$, the subscript $n$ indicates that the subtrees have $n$ generations, as opposed to $n+1$ for the full tree. Averaging over the bond energies, each drawn independently from the distribution $\rho(\epsilon)$ leads to the relation in Eq. (\ref{rec_DPRM}).

To analyze the traveling front solutions of a non-linear equation of the type (\ref{rec_DPRM}) one focuses on the left tail of $P_n(x)$, for $x \to -\infty$, where $1-P_n(x) \ll 1$. In this regime, where Eq. (\ref{rec_DPRM}) can be linearized, one looks for solution of the form $1-P_n(x) \approx \e^{-\lambda(x-v(\lambda)n)}$ which leads to the following dispersion spectrum \cite{Satya:00}
\begin{eqnarray}\label{dispersion}
v(\lambda) = - \frac{1}{\lambda} \ln \left[ 2 \int\, \rho(\epsilon)  \e^{-\lambda\,\epsilon} \,d\epsilon\right] \;.
\end{eqnarray}
For a generic distribution $\rho(\epsilon)$, the function $v(\lambda)$ admits a unique maximum at $\lambda^*$. The ``velocity selection principle'' states that the maximum velocity $v(\lambda^*)$ will be selected by the front \cite{saarloos}. One can then compute the two leading terms of $\langle E_{\rm min} \rangle$ in the large $n$ limit, which generically read 
\begin{eqnarray}\label{Emin_av}
\langle E_{\rm min} \rangle \approx v(\lambda^*) \,n\, + \frac{3}{2\lambda^*} \, \ln n \;.
\end{eqnarray}

This model of a directed polymer on a tree at zero temperature is also related to the so-called binary search tree (BST) problem in computer science.
The study of BST is an important problem in computer science, with many applications related to sorting and search of data. For instance, when
one stores data files in one's computer, usually one makes directories, sub-directories,  sub-sub-directories, etc, which has a natural tree structure.
The general idea is that, if one searches for a particular data, it is much more efficient to search if the data is stored on a tree. Typically, if the number of data
elements (e.g. the number of files) stored on the tree is of size $N$, then a binary search takes a time of order $O(\ln N)$, which is much smaller compared
to $O(N)$ for linear search. In fact, the search time is measured by the height $H_N$ of the BST, which is the maximal depth of the nodes occupied on the tree with $N$ occupied nodes \cite{Satya:02}. It turns out that there is a direct mapping between the BST and the directed polymer on the Cayley tree \cite{Satya:02}. The cumulative height distribution $P(H_N<n)$, for random data, in the BST problem is precisely related to the distribution of the minimum (ground state) energy of the directed problem discussed above. This relation in Eq. (\ref{Emin_av}), in the context of the BST, translates into a result for the average height \cite{Satya:02}
\begin{eqnarray}\label{mean_height}
\langle H_N \rangle \approx \frac{1}{v(\lambda^*)} \ln N - \frac{3}{2 \lambda^* v(\lambda^*)} \ln \ln N~,
\end{eqnarray}
where $v(\lambda)$ is defined in Eq. (\ref{dispersion}) above and $\lambda^*$ is the unique maximum of $v(\lambda)$. Similar traveling front solutions have also been found for other hierarchically correlated random variables, such as in a class of fragmentation models, aggregation dynamics of growing random trees, or in simple models of dynamics of efficiency (for a review see \cite{MK03} and \cite{Satya:05}). Interestingly,
although the dispersion spectrum $v(\lambda)$, and consequently $\lambda^*$ as well, are non-universal (i.e.,
they will differ from one hierarchically correlated model to another), the coefficient $3/2$ in front of the logarithmic correction in (\ref{Emin_av}) turns out
to be universal. 

For the directed polymer on the Cayley tree, it is also possible to study the full PDF of $E_{\min}$ in the limit of large $n$ \cite{Dean:01}. 
If the variables ${E_{\rm path}}_i$'s were IID random variables [see e.g. Eq. (\ref{def_Gumbel})], the PDF of ${E_{\min}}$, properly centred and scaled, would be given (for an unbounded distribution $\rho(\epsilon)$) by a Gumbel law $F'_2(-z) = \e^{z-\e^z}$ (note that we are considering here the minimum, at variance with the result in Eq. (\ref{def_Gumbel}) that concerns the maximum, which explains the `$-z$' in the argument). Instead of this, because the energy variables are actually hierarchically correlated, it was found that, for unbounded distribution $\rho(\epsilon)$, the limiting distribution is different from a Gumbel law. In particular, its right tail is non-universal and depends explicitly on $\rho(\epsilon)$. For instance, it was found~\cite{Dean:01} that for $\rho(\epsilon) = \e^{-|\epsilon|}/2$, the limiting distribution has a simple exponential tail $\sim \e^{-z}$ for $z \to +\infty$, which is quite different from the super-exponential tail of the Gumbel law.

These behaviors found for hierarchically correlated are reminiscent of the results obtained for {\it logarithmically} correlated 
random variables. For concreteness, let us indeed consider a set of $N$ Gaussian random variables $V_1, V_2, \ldots, V_i, \ldots, V_N$ where the subscript $i$
refers to a site index, which have logarithmic correlations, i.e.
\begin{eqnarray}\label{log_correlation}
\langle (V_i - V_j)^2\rangle \approx 4 \sigma \ln |i-j| \;, \; {\rm for} \quad \quad  1 \ll |i-j| \ll N  \;,
\end{eqnarray} 
where $\sigma > 0$, and denote $V_{\min} = \min \{V_1, V_2, \ldots, V_N \}$. This model was first studied by Carpentier and Le Doussal~\cite{CLD2001}, who found that the average value $\langle V_{\min}\rangle$ behaves indeed as described above in Eq.~(\ref{min_dp}) for hierarchically correlated random variables (with the substitution $n \to \ln N$). In addition, they showed that the {\it left} tail of the limiting distribution of $V_{\min}$ (properly shifted and scaled) behaves as $|z| \, \e^{z}$ as $z \to - \infty$, which is indeed different from the left tail of the Gumbel law $\sim \e^{z}$. One can further consider the simple model of a single particle in the disordered potential $V_i$ at site $i$, with $i=0,1, \cdots, N$ at thermal equilibrium at inverse temperature $\beta$. The probability $p_i$ to find the particle at site $i$ is thus given by the Boltzmann weight
\begin{eqnarray} \label{Boltzmann_log}
p_i = \frac{1}{Z_N} \, \e^{-\beta {V_i}} \quad \quad \;, \quad \quad Z_N = \sum_{i=1}^N \e^{-\beta{V_i}} \;,
\end{eqnarray}
where the $V_i$'s are Gaussian random variables with logarithmic correlations as in (\ref{log_correlation}). Using Renormalization Group techniques, it was shown that this model~(\ref{Boltzmann_log}) exhibits an interesting freezing phenomenon at a finite inverse temperature $\beta_g$, which is reminiscent of the transition found in Derrida's Random Energy Model (REM) \cite{Derrida:81} and its generalisations \cite{Derrida:85,DG86}. Indeed, at high temperature $\beta < \beta_g$, the particle is delocalised over the whole sample while for $\beta > \beta_g$ the Boltzmann weight is concentrated on the few local minima of the disordered potential \cite{CLD2001}. Such a freezing transition was later on confirmed by the exact solution of several models of logarithmically correlated random variables such as the equilibrium properties of single particle in random Gaussian high-dimensional landscape~\cite{FB07, FB08, FS07} or a circular variant of the REM with logarithmic correlations \cite{FB08}. For these different models, the full limiting distribution of the minimum is not universal, i.e. it depends on the details of the model, but the left tail turns out to be universal, and behaves as $|z| \, \e^{z}$ as $z \to - \infty$ \cite{CLD2001}. Since then, these models have generated a lot of interest, both in physics and in mathematics, because of their relations with the extremes of the two-dimensional Gaussian Free Field ($2d$-GFF) \cite{FLR09,BZ12,BDZ16}. More recently, it was suggested that the same freezing transition also governs the extreme values taken by the characteristic polynomials of random matrices and the Riemann zeta function \cite{FHK12, FK2014}. This, in turn, has generated a recent interest in mathematics \cite{ABB17,ABH17,PZ18}, in connection with random matrices and branching processes.

\subsubsection{Extreme statistics in stochastic resetting systems}

We already discussed in the previous sub-section how the extreme value statistics plays an important role in binary search tree problems where the
random data is stored on a binary tree. EVS has also found applications in other stochastic search processes, such as the diffusive search 
of a target by an animal. In particular, a recently proposed search model namely stochastic resetting \cite{restart1,restart2} has been quite successful to explain animal foraging \cite{search1,search2}, bio-molecular search \cite{search3} and chemical kinetics \cite{search4} (for a recent review see \cite{EMS19}). Resetting is a stochastic process where an underlying process is intermittently reset to a preferred
configuration with a given probability \cite{restart1,restart2}. This simple yet pivotal setup has become an emerging and overarching topic in interdisciplinary sciences since restarting a complex process again and again can 
expedite the completion of such process and thus can be engineered as a useful tool.
To understand this, let us consider a simple 1D Brownian motion (starting from $x_0$) in the presence of a target at the origin. The motion is reset to its initial configuration
with a rate $r$, and one is interested in the mean first passage time to the target. Although the mean first passage time is divergent when $r=0$ \cite{redner}, a finite resetting rate renders the mean first passage time finite and thus 
facilitates a diffusion-mediated-search which is otherwise detrimental \cite{restart1,restart2}. 

\begin{figure}[t]
\includegraphics[width = 0.6\linewidth]{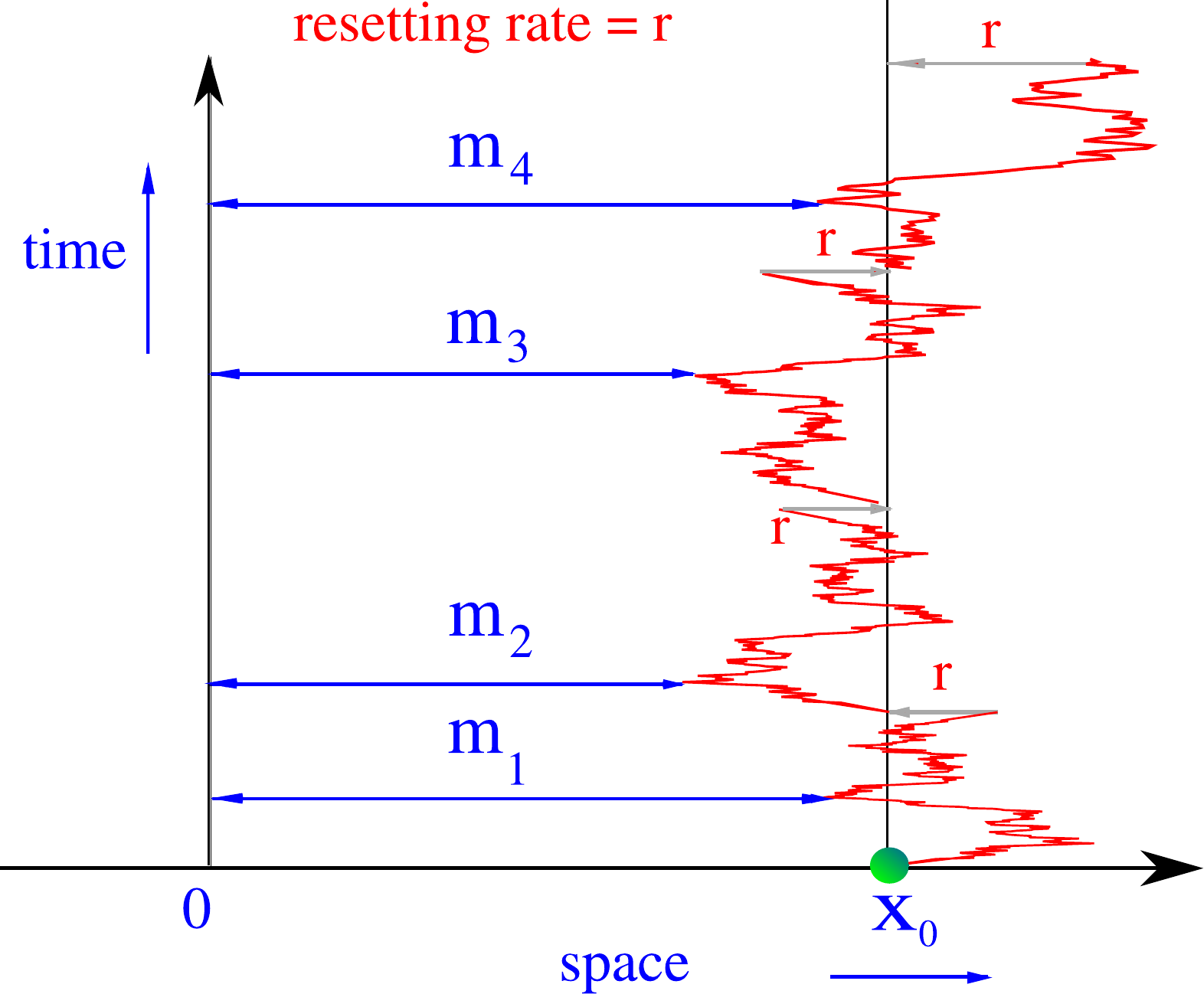}
\caption{Illustration of a single trajectory of the one-dimensional diffusion process with stochastic resetting: the searcher performs normal diffusion that gets interrupted randomly at rate $r$ whereby the searcher restarts the search from its initial position $x_0 > 0$. The resetting events are shown by the grey arrows. 
We assume that there is an immobile target at the origin $0$. Within each block $i =1, 2, \cdots, N$ (here $N=4$) between two successive resetting events we denote by $m_i$ the minimum of the Brownian
trajectory within this block. The $m_i$'s are IID random variables since the process is renewed after each resetting. As explained in the text, these variables $m_i$'s are useful to compute the survival probability of the target at the origin using~EVS.}\label{Fig_reset}
\end{figure}
To see how EVS appears in this stochastic resetting problem, we consider the one-dimensional diffusion with stochastic resetting at a 
constant rate $r$ \cite{restart1,restart2}. The searcher starts at a position $x_0>0$ and we assume that there is an immobile target at the origin.
The searcher performs normal diffusion with diffusion constant $D$ that gets interrupted at rate $r$ after which the searcher starts again from its initial position $x_0$ (see Fig. \ref{Fig_reset}). Let $Q_r(x_0,t)$ denote the survival probability $Q_r(x_0,t)$ of the target, i.e., the probability that the searcher has not found the target up to time $t$. This quantity has been exactly computed using a backward Fokker-Planck as well as a renewal approach \cite{restart1} and a simple interpretation of this result in terms of EVS was provided \cite{restart1}. The argument goes as follows. We consider this stochastic process up to a total observation time $t$. Since the mean time between reset event is typically $1/r$, the mean number of intervals between resets is of order $N = r \, t$. The target remains un-hit by the searcher up to time $t$ if the global minimum of this process, starting at $x_0$, remains positive (since the target is at the origin and $x_0>0$). To compute the global minimum, we can think of the total interval $[0,t]$ consisting of $N = r\,t$ uncorrelated blocks (see Fig. \ref{Fig_reset}). 
For each block $i$, there is a minimum $m_i$ and the $m_i$'s are uncorrelated since the process is renewed after each resetting. The global minimum of the process is also the minimum of the $m_i$'s, $i=1,2, \cdots, N = r\,t$. The cumulative PDF of $m_i$, i.e. ${\rm Prob}(m_i>m)$, is given by the Brownian motion result (since inside a block the particle performs normal diffusion), ${\rm Prob}(m_i>m) = {\rm } = {\rm 
erf}\left(\frac{|M-x_0|}{\sqrt{4D\tau_i}}\right)$ [see for instance Eq. (\ref{CDF})], where $\tau_i$ is the duration of the $i$-th block. Thus the PDF of $m_i$ is given by $P(m_i)=\e^{-(m_i-x_0)^2/{4D\tau_i}}/\sqrt{\pi D\tau_i}$ with $m_i\le x_0$. Since these time intervals between resets are taken from an exponential distribution with mean $1/r$, one has  $P(\tau_i)=r 
\e^{-r\tau_i}$. Averaging over $\tau_i$, one gets the effective distribution of the minimum of the $i$-th block as
\begin{eqnarray}\label{eff_PDF}
P_{\rm eff}(m_i)= r\int_{0}^t d\tau_i e^{-r \tau_i} \frac{\e^{-(m_i-x_0)^2/{4D\tau_i}}}{\sqrt{\pi D\tau_i}} \underset{t \to \infty}{\longrightarrow} \alpha_0 \e^{-\alpha_0(x_0-m_i)} \;, \;\; m_i \leq x_0 \;.
\end{eqnarray}
where $\alpha_0=\sqrt{\frac{r}{D}}$ is the typical inverse length traversed by the Brownian motion between two resetting events. Therefore,
the probability
that the minimum of all $N$ intervals stays above $0$ is simply given by
$[\int_0^{x_0} P_{\rm eff}(m)\, dm]^N$. Substituting the expression for $P_{\rm eff}(m)$ from Eq. (\ref{eff_PDF}), setting $N = r\,t$, one finds that, 
in terms of the rescaled length $z=\alpha_0 x_0$ the survival probability converges for large $z$ and large $t$ to the Gumbel form
\bea
Q_r(x_0,t) \approx \exp\left[-rt 
\exp(-z)\right].
\label{Q_r-1}
\eea 
Therefore, this constitues a nice application of the EVS for weakly correlated variables discussed in Section \ref{Sec:weakly}. The theory of EVS can also be useful to understand the first-passage properties of other stochastic processes with resetting, going beyond the Brownian motion. This includes, for example, L\'evy flights \cite{search1} and L\'evy walks \cite{KG15} under resetting, branching processes under restart \cite{restart-branching, restart3}, active run-and-tumble particles with resetting \cite{EV2018}, etc. We do not give further details here and refer the reader to the relevant literature cited above.

\subsubsection{Extreme statistics for fluctuating interfaces in one-dimension: an example of a strongly correlated random variable}

Another simple model where the extreme value statistics has been studied concerned the height fluctuations of a (1+1)-dimensional 
interface. As we will see below, this is an example where the relevant degrees of freedom are strongly correlated. 
The most well studied model of a such fluctuating interface is the so called Kardar-Parisi-Zhang (KPZ)
equation that describes the time evolution of the height $H(x,t)$ of an interface 
growing over a linear substrate of size $L$ via the stochastic partial
differential equation \cite{KPZ86}
\bea
\f{\partial H}{\partial t}=\f{\partial^{2}H}{\partial x^{2}}+\lambda 
\left(\f{\partial H}{\partial x}\right)^{2}+\eta(x,t)~,
\eea
where $\eta(x,t)$ is a Gaussian white noise with zero mean and a 
correlator $\langle \eta(x,t)\eta(x^{\prime},t^{\prime}) \rangle=2\delta(x-x^{\prime})\delta(t-t^{\prime})$. 
For $\lambda=0$, the equation becomes linear and 
is known as the Edwards-Wilkinson (EW) equation. 
For reviews on fluctuating interfaces, see e.g.~\cite{HZ95,Krug97,Spohn_Houches,HT15}.
The height is usually measured 
relatively to the spatially averaged height i.e.
\bea
h(x,t)=H(x,t)-\f{1}{L}\int_{0}^{L}H(y,t)dy~, \\
\text{with~}\int_{0}^{L}h(x,t)dx=0 \;.
\eea
It can be shown that the joint PDF of the relative height field
$P(\{h\},t)$ reaches a steady state as $t \to \infty$ in a finite system of size $L$.
Also the height variables are strongly correlated in the stationary state.
Again in the context of the EVS, a quantity that has been studied during the last few years 
is the PDF of the maximum relative height in the stationary state, i.e. $P(h_{m},L)$
where
\bea
h_{m} = \lim_{t \to \infty} \text{max}_x [{h(x, t)}, 0 \leq x \leq L] .
\eea
This is an important physical quantity that measures the extreme fluctuations 
of the interface heights~\cite{Raychowdhury,Satya:04,airy}. We assume
that initially the height profile is flat. 
As time evolves, the heights of the interfaces at different spatial points
get more and more correlated. The correlation length typically grows as
$\xi\sim t^{1/z}$ where $z$ is the dynamical exponent ($z=3/2$ for KPZ and
$z=2$ for EW interfaces). For $t \ll L^{z}$, the interface is in the `growing' regime
where again the height variables are weakly correlated since $\xi\sim t^{1/z} \ll L$.
In contrast, for $t \gg  L^z$, the system approaches a `stationary' regime
where the correlation length $\xi$ approaches the system size and hence
the heights become strongly correlated variables.

Following our general argument for weakly correlated variables (see Section \ref{Sec:weakly}), we would then
expect that in the growing regime the maximal relative height, appropriately centred
and scaled, should have the Gumbel distribution.
In contrast, in the stationary regime, the height variables are strongly
correlated and the maximal relative height $h_m$ should have a different
distribution. This distribution was first studied numerically in ~\cite{Raychowdhury}
and then it was computed analytically in Refs.~\cite{Satya:04,airy}. This then
presents one of the rare solvable cases for the EVS of strongly correlated
random variables. Below, we briefly outline the derivation of this distribution. 

The joint PDF of the relative heights in the stationary state can be written, taken into account the different contraints~\cite{Satya:04,airy}
\bea
P_{st}[\{h\}]=C(L)\, \e^{-\f{1}{2}\int_{0}^{L}(\partial_{x}h)^{2}~dx} \times \delta \Big[h(0)-h(L)\Big] \times \delta\Big[\int_{0}^{L}h(x,t)dx\Big]~,
\label{stat_measure}
\eea
where $C(L)=\sqrt{2\pi L^{3}}$ is the normalization constant and can be obtained integrating over all 
the heights. Note that this stationary measure of the relative heights 
is independent of the coefficient $\lambda$ of the nonlinear term in the KPZ equation, implying
that the stationary measure of the KPZ and the EW interface is the same
in $(1+1)$-dimension (note however that this a special property of the $1+1$-dimensional case, which breaks
down for $(d+1)$-dimensional KPZ interfaces). The stationary measure indicates that the interface behaves locally as a Brownian motion 
in space~\cite{HZ95,Krug97,Spohn_Houches}. 
For an interface with periodic boundary condition, one would then
have a Brownian bridge in space.
However, 
it turns out that the constraint $\int_0^L h(x,t)\, dx=0$ (the zero mode
being identically zero), as shown explicitly
by the delta function in  Eq. (\ref{stat_measure}), plays an important role for
the statistics of the maximal relative height~\cite{Satya:04}. It shows actually that
the stationary measure of the relative heights actually corresponds to
a Brownian bridge, but with a global constraint that the area under
the bridge is strictly zero~\cite{Satya:04,airy}. This fact plays a
crucial role for the extreme statistics of relative heights~\cite{Satya:04,airy}.

We define the cumulative distribution of the maximum relative height
$Q(z,L)=\text{Prob}[h_{m}\leq z]$. The PDF of the maximum relative height is 
then $P(z,L)=Q^{\prime}(z,L)$. 
Clearly $Q(z,L)$ is also the probability that the heights
at all points in $[0, L]$ are less than $z$ and can be formally
written in terms of the path integral~\cite{Satya:04,airy}
\bea
Q(z,L)=C(L)\int_{-\infty}^{z}du\int_{h(0)=u}^{h(L)=u}\mathcal{D}h(x)\e^{-\f{1}{2}\int_{0}^{L}(\partial_{x}h)^{2}~dx} \nn \\
\times  \delta\Big[\int_{0}^{L}h(x,t)dx\Big]I(z,L)~,
\eea
where $I(z,L)=\prod_{x=0}^{L}\theta[z-h(x)]$ is an indicator function 
which is $1$ if all the heights are less than $z$ and zero
otherwise. Using path integral techniques (for details see ~\cite{Satya:04,airy}). 
it was found that the PDF of $h_m$ takes the scaling form for all~$L$
\bea
P(h_{m},L)=\f{1}{\sqrt{L}}f\left(\f{h_{m}}{\sqrt{L}}\right) \;,
\eea
where the scaling function can be computed explicitly
as~\cite{Satya:04,airy}
\bea
f(x)=\f{2\sqrt{6}}{x^{10/3}}\sum_{k=1}^{\infty}\e^{-\f{b_{k}}{x^{2}}}\,b_{k}^{2/3}\,U\left(
-\f{5}{6},\f{4}{3},\f{b_{k}}{x^{2}}\right) \;,
\eea
where $U(a,b,y)$ is the confluent hypergeometric function 
and $b_{k}=\f{2}{27}\alpha_{k}^{3}$, where $\alpha_k$'s are
the absolute values of the zeros of the Airy function: ${\rm Ai}(-\alpha_k)=0$.
It is easy to obtain the small $x$ behavior of $x$ since 
only the $k=1$ term dominates as $x \to 0$.
Using $U(a,b,y)\sim y^{-a}$ for large $y$, we get as $x \to 0$,
\bea
f(x)\approx \f{8}{81}\alpha_{1}^{9/2}\,x^{-5}\,\exp\left[-\f{2 \alpha_{1}^{3}}{27x^{2}}\right] \;, \; x \to 0 \;.
\eea
The asymptotic behavior of $f(x)$ at large $x$ can be obtained as \cite{Satya:04,airy,JL07,RBKS11}
\bea
f(x)\approx 72 \sqrt{\frac{6}{\pi}}\, x^2 \,\e^{-6 x^{2}}  \;, \; x \to \infty \;.
\eea
It turns out, rather interestingly, that this same function has appeared before
in several different problems in computer science and probability theory and
is known in the literature as the Airy distribution function. In particular, it describes the distribution of the area
under a Brownian excursion over the unit time interval (for a review on
this function, its occurence in different contexts and its generalisations see Refs.~\cite{airy,Maj05,Jan07}
and references therein). 

The path integral technique mentioned above to compute the
maximal relative height distribution of the EW/KPZ stationary interfaces
have subsequently been generalised to more complex interfaces~\cite{SM06,GMOR07,Burkhardt,RS09,RBKS11,KMS07, KM14}.

\subsubsection{Extreme statistics in random matrix theory}

Another beautiful solvable example of the extremal statistics of strongly correlated variables
can be found in random matrices~\cite{TW94,TW96} (see~\cite{Satya:14} for a recent review). Let us consider a $N\times N$ Gaussian random matrices 
with real symmetric, complex Hermitian,
or quaternionic self-dual entries
$X_{i,j}$ distributed via the joint Gaussian law \cite{mehta,forrester}
\bea
\text{Pr}[\{X_{i,j}\}]\propto \exp[-\f{\beta}{2}N\text{Tr}(X^{2})] \;, \label{Gauss_RMT}
\eea
where $\beta$ is the so called Dyson index. The distribution is invariant respectively
under orthogonal, unitary and symplectic transformations giving rise
to the three classical ensembles: Gaussian orthogonal ensemble (GOE),
Gaussian unitary ensemble (GUE) and Gaussian symplectic ensemble (GSE).
The quantized values of $\beta$ are respectively $\beta=1~$(GOE),
$\beta=2~$(GUE) and $\beta=4~$(GSE). The eigenvalues and eigenvectors turn out to be independent and their joint distribution thus factorizes. Integrating out the eigenvectors
we focus here only on the statistics of $N$ eigenvalues $\lambda_{1}, \lambda_{2},...,~\lambda_{N}$ which are all real. The joint PDF of these eigenvalues
is given by the classical result \cite{mehta,forrester}
\bea\label{joint_GUE}
P_{\text{joint}}(\lambda_{1}, \cdots, \lambda_{N})=B_{N}(\beta)\exp \bigg[ -
\f{\beta}{2}N\sum_{i=1}^{N} \lambda_{i}^{2} \bigg] \prod_{i<j}|\lambda_{i}-\lambda_{j}|^{\beta},
\eea
where $B_{N}(\beta)$ is the normalization constant. For convenience, we rewrite the statistical weight as
\bea\label{joint_pdf}
P_{\text{joint}}(\lambda_{1}, \cdots, \lambda_{N})=B_{N}(\beta)
\exp \Bigg[ -\beta \Big( \f{N}{2}\sum_{i=1}^{N} \lambda_{i}^{2}-\f{1}{2} \sum_{i \neq j}\ln |\lambda_{i}-\lambda_{j}| \Big) \Bigg]~.
\eea
Hence, this joint law can be interpreted as a Gibbs-Boltzmann
weight \cite{Dys62} 
$P_{\rm joint}(\{\lambda_i\})\propto \exp\left[-\beta\,
E\left(\{\lambda_i\}\right)\right]$,
of an interacting gas of charged particles on a line where $\lambda_i$
denotes
the position of the $i$-th charge and $\beta$ plays the role of the
inverse temperature. The energy $E\left(\{\lambda_i\}\right)$ has
two parts: each pair of charges repel
each other via a $2d$ Coulomb (logarithmic)
repulsion (even though the charges are confined on the $1d$
real line)
and
each charge is subject to an external confining parabolic potential.
Note that while $\beta = 1$, $2$ and $4$ correspond to the three
classical rotationally invariant Gaussian ensembles described by the measure in Eq. (\ref{Gauss_RMT}), it is
possible to associate a   
matrix model
to~(\ref{joint_pdf}) for any value of $\beta > 0$ (namely tridiagonal
random matrices introduced in \cite{DE02}). These ensembles defined as in Eq. (\ref{joint_GUE}) for generic $\beta$ are sometimes called
the ``Gaussian $\beta$-ensembles''.  Here we focus on
the largest eigenvalue $\lmax = \max_{1 \leq i \leq N} \lambda_i$: what
can be said about its fluctuations, in particular when $N$ is large ?
This
is a nontrivial question as the interaction term, $\propto
|\lambda_i - \lambda_j|^\beta$, renders the classical results
of extreme value statistics for
IID random variables discussed in Section \ref{sec:IID} inapplicable.

The two terms in the energy of the Coulomb gas in (\ref{joint_pdf}), the
pairwise Coulomb repulsion and the external harmonic potential, compete
with each other. While the former tends to spread the charges apart, the  
later tends to confine the charges near the origin. The average density of charges is given by
\bea
\rho_{N}(\lambda)=\f{1}{N}\Big \langle
\sum_{i=1}^{N}\delta(\lambda-\lambda_{i}) \Big \rangle~,
\eea
where the angular brackets denote an average with respect to the joint
PDF in Eq. (\ref{joint_pdf}). For such Gaussian matrices (\ref{joint_pdf}), it is well
known \cite{mehta, forrester, Wig51} that as $N\to \infty$,
the average density approaches an $N$-independent limiting form which
has a semi-circular shape
on the compact support $[-\sqrt{2}, + \sqrt{2}]$
\bea
\lim_{N\to \infty} \rho_{N}(\lambda)=\tilde{\rho}_{sc}(\lambda)=\f{1}{\pi}\sqrt{2-\lambda^{2}} \;, \label{Wigner_sc}
\eea
where $\tilde{\rho}_{sc}(\lambda)$ is called the Wigner semi-circular law. Hence our first observation is 
that the maximum eigenvalue resides near the upper edge of the Wigner semi-circle:
\bea
\lim_{N\to \infty} \langle \lambda_{\text{max}} \rangle=\sqrt{2} \;.
\eea
However, for large but finite $N$, $\lambda_{\text{max}}$ will
fluctuate from sample to sample and the goal is to compute its cumulative distribution  
\bea
Q_{N}(w)=\text{Prob}[\lambda_{\text{max}}<w]~, \label{def_QN_RMT}
\eea
which can be written as a ratio of two partition functions
\bea \label{ratio}
Q_{N}(w)&=&\f{Z_{N}(w)}{Z_{N}(w\to \infty)}, \\
Z_{N}(w)&=&\int_{-\infty}^{w}d\lambda_{1}...\int_{-\infty}^{w}d\lambda_{N}\exp \Bigg[ 
-\beta \Big( \f{N}{2}\sum_{i=1}^{N} \lambda_{i}^{2}-\f{1}{2} \sum_{i \neq j}\ln |\lambda_{i}-\lambda_{j}| \Big) \Bigg]  \;.
\eea
The partition function $Z_N(w)$ in the numerator describes a $2d$ Coulomb gas, confined 
on a $1d$ line and subject to a harmonic potential, as in Eq. (\ref{joint_pdf}), but now in the presence of an impenetrable {\it hard wall} at $w$.
The study of this ratio of two partition functions reveals the existence of 
two distinct scales corresponding to (i) {\it typical} fluctuations of the 
top eigenvalue, where $\lambda_{\text{max}}=\mathcal{O}(N^{-2/3})$
and (ii) to {\it atypical} large fluctuations, where $\lambda_{\text{max}}=\mathcal{O}(1)$. 
\begin{figure}[t]
\includegraphics[width = 0.6\linewidth]{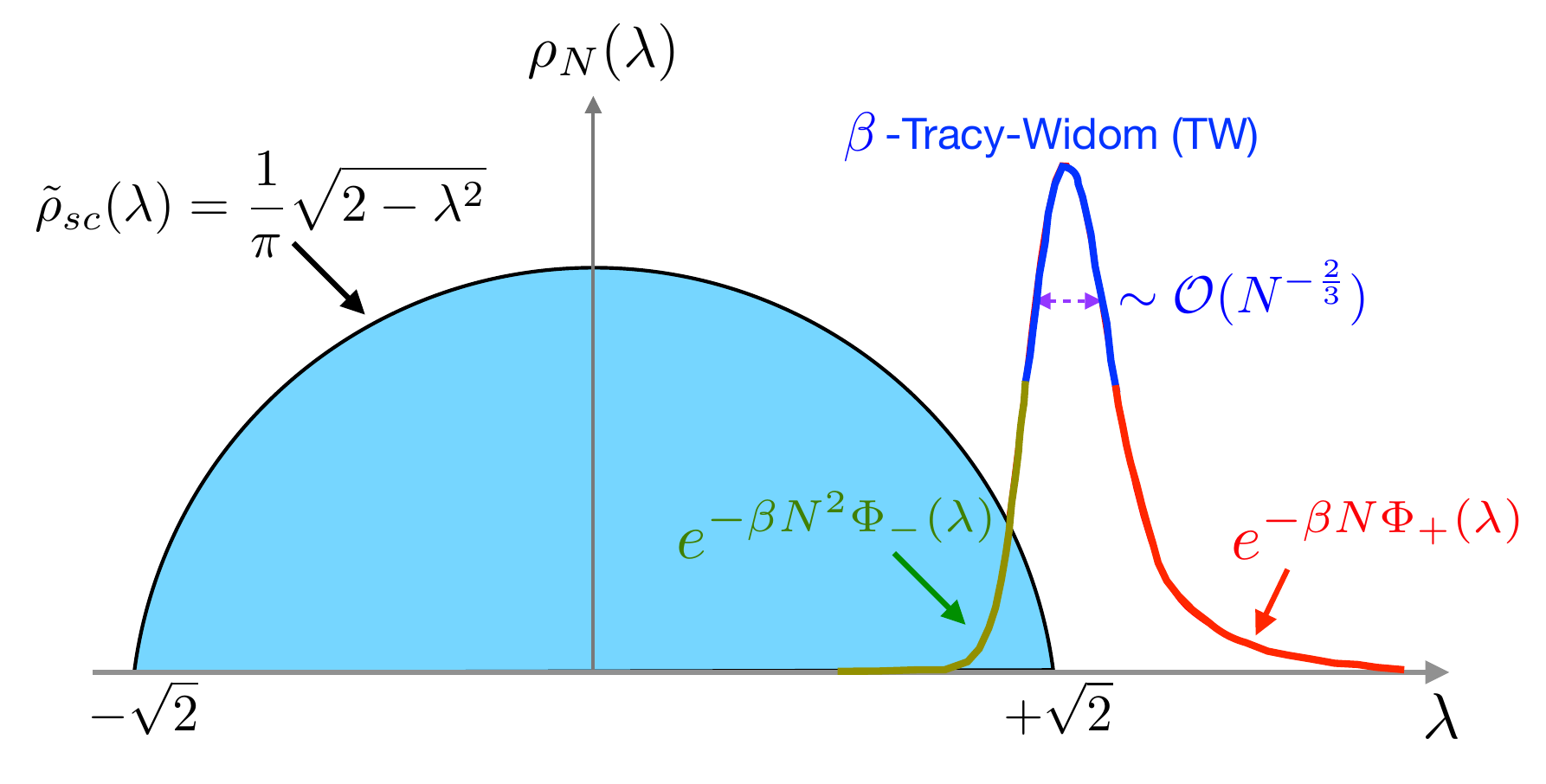}
\caption{Sketch of the PDF of $\lambda_{\max}$ with a peak around the right edge of the Wigner semi-circle given in Eq. (\ref{Wigner_sc}) at $\langle \lambda_{\max}\rangle = \sqrt{2}$. The {\it typical} fluctuations around the mean are described by the $\beta$-Tracy-Widom distribution (blue), while the large deviations of order ${\cal O}(1)$ to the left and the right of the mean are described by the left (green) and right (red) large deviation tails [see Eq.~(\ref{summary})].}\label{Plot_TW}
\end{figure}

{More precisely, in the regime (i) of typical fluctuations,
it can be shown that as $N \to \infty$
\bea
\lambda_{\text{max}} \approx \sqrt{2}+\f{1}{\sqrt{2}}N^{-2/3}\chi_{\beta}~,
\eea
where $\chi_{\beta}$ is an $N$-independent random variable. Its cumulative distribution,
$\mathcal{F}_{\beta}=\text{Prob}[\chi_{\beta}\leq x],$ is known as the 
$\beta$-Tracy-Widom (TW) distribution. For $\beta = 1, 2, 4$, ${\cal F}_{\beta}$ can be written explicitly in terms of
a special solution of a Painlev\'e equation~\cite{TW94,TW96} and for $\beta = 6$ in terms of an additional Painlev{\'e} transcendent \cite{Rum16, GIKM16}. For other values of $\beta$, it can be shown that ${\cal F}_\beta$ describes the fluctuations of the ground state of a one-dimensional random Schr\"odinger operator, called the ``stochastic Airy operator'' \cite{ES07,RRV11}. For arbitrary $\beta>0$, it can be shown that the PDF ${\cal F}'_\beta(x)$ has asymmetric 
non-Gaussian tails~\cite{TW94,TW96},
\bea\label{tail_TW}
\mathcal{F}_{\beta}^{\prime}(x) \approx \left\{
     \begin{array} {rl}
     {\rm \exp}{~\Big[-\f{\beta}{24}|x|^{3}\Big]} \;, & ~~~x \to  -\infty\\
         \\
     {\rm \exp}{~\Big[-\f{2\beta}{3}x^{3/2}\Big]} \;. & ~~~x \to  +\infty \\
     \end{array} \right.
\eea
We refer the reader to the Appendix \ref{Sec:TW} as well as to Refs. \cite{BEMN11,BBD08,DIK08} and \cite{BN12,DV13} for more precise asymptotics of the left and right tail of the $\beta$-TW distribution respectively.} {Quite remarkably, the same TW 
distributions (in particular for $\beta = 1, 2$) have emerged in a number of a priori unrelated problems 
\cite{maj} such as the longest increasing subsequence of random 
permutations \cite{baik}, directed polymers \cite{johann,poli} and growth 
models \cite{growth1,growth2,growth3,growth4} in the Kardar-Parisi-Zhang (KPZ) universality class in 
$(1+1)$ 
dimensions as well as for the continuum $(1+1)$-dimensional KPZ 
equation~\cite{SS10,CLR10,DOT10,ACQ11}, 
sequence alignment problems 
\cite{sequence}, mesoscopic 
fluctuations in quantum dots \cite{dots1,dots2,dots3,dots4}, height fluctuations of 
non-intersecting Brownian motions over a fixed time interval
\cite{FMS11,Lie12}, height fluctuations of non-intersecting interfaces
in the presence of a long-range interaction induced by a 
substrate~\cite{NM_interface} or more recently in the context of trapped fermions \cite{DDMS15, DDMS16} (see below in Section \ref{Sec:fermions}), as well as 
in finance 
\cite{biroli}. Remarkably, the TW distributions have also been observed in experiments on nematic liquid crystals \cite{takeuchi1,takeuchi2,takeuchi3} (for 
$\beta = 1,2$) and in experiments involving coupled fibre lasers 
\cite{davidson} (for $\beta = 1$).}

{While the $\beta$-TW distribution describes the typical fluctuations 
of $\lambda_{\text{max}}$ around
its mean $\langle \lambda_{\text{max}} \rangle=\sqrt{2}$ on a small 
scale of $\mathcal{O}(N^{-2/3})$, it does not describe atypically
the large fluctuations, e.g. of order $\mathcal{O}(1)$ around $\langle \lambda_{\text{max}}\rangle$.
The probability of atypically large
fluctuations, to leading order for large $N$, is described by two 
large deviations (or rate)
functions $\Phi_{-}(w)$ (for fluctuations to the left of the mean) 
and $\Phi_{+}(w)$ (for fluctuations to
the right of the mean). These different regimes for the cumulative distribution $Q_N(w)$ of $\lambda_{\max}$ in Eq. (\ref{def_QN_RMT}) can be summarised, at leading order for large $N$, as follows
\bea\label{summary_CDF}
Q_N(w) \approx \left\{
     \begin{array} {rl}
     {\rm \exp}{~\Big[-\beta N^{2}\Phi_{-}}(w)\Big], & ~~~w<\sqrt{2}~\text{and}~|w-\sqrt{2}|\sim \mathcal{O}(1)\\
         \\
      {\mathcal{F}_{\beta}\Big[\sqrt{2}N^{2/3}}(w-\sqrt{2})\Big], & ~~~|w-\sqrt{2}|\sim \mathcal{O}(N^{-2/3})\\
          \\
   1-  {\rm \exp}{~\Big[-\beta N\Phi_{+}}(w)\Big], & ~~~w>\sqrt{2}~\text{and}~|w-\sqrt{2}|\sim \mathcal{O}(1) \;. \\
     \end{array} \right.
\eea
Equivalently, the PDF of $\lambda_{\max}$ obtained from $P(\lambda_{\text{max}}=w,~N) = d Q_N(\lambda)/d\lambda$ reads, at leading order for large $N$ (see also Fig. \ref{Plot_TW})
\bea\label{summary}
P(\lambda_{\text{max}}=w,~N) \approx \left\{
     \begin{array} {rl}
     {\rm \exp}{~\Big[-\beta N^{2}\Phi_{-}}(w)\Big], & ~~~w<\sqrt{2}~\text{and}~|w-\sqrt{2}|\sim \mathcal{O}(1)\\
         \\
      {\sqrt{2}N^{2/3}\mathcal{F}_{\beta}^{\prime}\Big[\sqrt{2}N^{2/3}}(w-\sqrt{2})\Big], & ~~~|w-\sqrt{2}|\sim \mathcal{O}(N^{-2/3})\\
          \\
     {\rm \exp}{~\Big[-\beta N\Phi_{+}}(w)\Big], & ~~~w>\sqrt{2}~\text{and}~|w-\sqrt{2}|\sim \mathcal{O}(1)\\
     \end{array} \right.
\eea
Note that in Eqs. (\ref{summary_CDF}) and (\ref{summary}), the symbol $\approx$ means a logarithmic equivalent. We refer to Refs. \cite{BEMN11} and \cite{NM11,For12,BN12} for the study of the sub-leading corrections to the large deviations forms (\ref{summary_CDF})-(\ref{summary}) both for the left tail and the right tail respectively. For the left tail, the rate function $\Phi_-(w)$ was first explicitly computed in Ref.~\cite{DM06,DM08}, using Coulomb gas techniques. In particular, when $w$ approaches the critical value $\sqrt{2}$ from below behaves as
\bea \label{phim_crit}
\Phi_{-}(w)\sim \f{1}{6\sqrt{2}}(\sqrt{2}-w)^{3},~~w\underset{w<\sqrt{2}}\longrightarrow~\sqrt{2}  \;.
\eea
For the right tail, the large deviation function $\Phi_+(w)$ was computed in Ref.~\cite{MV09}. A rigorous derivation (but only valid for $\beta =1$) of $\Phi_+(w)$ can be found in \cite{BDG01} (see also Ref. \cite{Fyo04} for yet another derivation of $\Phi_+(w)$ for $\beta = 1$). In particular, near $w = \sqrt{2}$ it behaves as
\bea \label{phip_crit}
\Phi_{+}(w)\sim \f{2^{7/4}}{3}(w-\sqrt{2})^{3/2},~~w\underset{w>\sqrt{2}}\longrightarrow~\sqrt{2}\;.
\eea
By inserting these asymptotic behaviors of the rate functions given by Eqs. (\ref{phim_crit}) and (\ref{phip_crit}) into Eq. (\ref{summary}), one can check that there is a smooth matching between the left and right large deviation tails and the tails of the central part described by the $\beta$-TW distribution (\ref{tail_TW}) -- see Ref. \cite{Satya:14} for more details. 
}

{Interestingly, the study of the large deviations of $\lambda_{\max}$ reveals the existence of a phase transition, separating the left and right tails. Indeed this corresponds to a thermodynamical phase transition for the free energy, $\propto \ln Q_N(w)$, of a Coulomb gas in the presence of a wall (\ref{ratio}) as the position of the wall crosses the value $w_c = \sqrt{2}$. Indeed, from Eq. (\ref{summary_CDF}), one has
\bea\label{transition}
\lim_{N \to \infty} - \frac{1}{N^2} \ln Q_N(w) = 
\begin{cases}
&\Phi_-(w) \;, \; w < \sqrt{2} \;, \\
&0 \;, \; w > \sqrt{2} \;.
\end{cases}
\eea
Since $\Phi_-(w) \propto (\sqrt{2} - w)^3$ [see Eq. (\ref{phim_crit})], the third derivative of the free energy of this Coulomb gas is discontinuous at $w_c = \sqrt{2}$: this indicates a {\it third order phase transition} \cite{NM11}. It turns out that this transition is very similar to the so called Gross-Witten-Wadia phase transition found in the 80's in the context of $2d$ lattice quantum chromodynamics~\cite{GW80,Wad80}. As discussed in further details in \cite{Satya:14}, a similar third-order transition occurs in various physical systems, including non-intersecting Brownian motions \cite{FMS11,SMCF13}, conductance fluctuations in mesoscopic physics \cite{VMB08,VMB10,DMTV11} and entanglement in a bipartite system \cite{NMV10,NMV11}. In Ref. \cite{Satya:14}, it was further argued that the universality of the TW distribution is inherited from the universality of phase transitions.}

Here, we focused on the Gaussian $\beta$-ensembles (\ref{joint_GUE}) but extreme value questions have also been studied for a wide variety of random matrix ensembles. This includes invariant ensembles, like the Laguerre-Wishart, the Jacobi or the Cauchy ensembles (for which all the eigenvalues are real) as well as non invariant ensembles, and in particular the so called Ginibre ensembles of RMT, for which the eigenvalues lie in the complex plane. In the latter case, the joint PDF of the complex eigenvalues can be identified with the Boltzmann weight of a two-dimensional Coulomb gas in the presence of an external quadratic confining potential, analogous to Eq. (\ref{joint_pdf}) but for complex eigenvalues $\lambda_i \to z_i$ where $z_i$'s are complex numbers. In these two-dimensional situations a natural extreme observable is the largest radius $r_{\max} = \max_{1 \leq i \leq N} |z_i|$~\cite{Rid2003, CP2014,CMV2016} (see also the discussion at the end of Section \ref{Sec:fermions} below). Further extensions have also been considered, either by studying various confining potentials (i.e., different from the quadratic well), which in some cases correspond to other interesting matrix models \cite{LGMS2018}, or by studying Coulomb gases in higher dimensions $d>2$ \cite{CFLV2018}. In the latter case many extreme value questions, like the distribution of the typical fluctuations of largest radius, remain open.

\subsubsection{Extreme statistics of trapped fermions}\label{Sec:fermions}

More recently EVS have been studied for non-interacting fermions in a trap. Because of the Pauli principle, the positions (or momenta) of a Fermi gas are strongly correlated variables even in the absence of interactions. Remarkably, the recent developments of so called ``Fermi quantum microscopes'' open the possibility to probe these correlations via a direct in situ imaging of the individual fermions, with a resolution comparable to the inter-particle spacing. The theoretical understanding of these spatio-temporal correlations in noninteracting fermions is therefore a challenging problem. Many of these experiments are performed in the presence of a trapping potential, which affects the spatial correlations in a non-trivial way, as it breaks the translational invariance of the system. The physics in the bulk near the trap center (where the fermions do not feel the curvature of the confining trap) can be understood using the traditional theories of quantum many-body systems such as the local density approximation (LDA) \cite{castin}. However, away from the trap center, the fermions start feeling the curvature induced by the confining trap. As a result the average density profile of the fermions vanishes beyond a certain distance from the trap center -- thus creating a sharp edge. Near this edge, the density is small (there are few fermions) and consequently, quantum and thermal fluctuations play a more dominant role than in the bulk. The importance of these fluctuations means that traditional theories such as LDA break down in this edge region \cite{KM98}. Recently, there has been some progress, using techniques from RMT and determinantal processes, to study analytically this edge region (for a short review see \cite{DLMS19}). In particular, a natural observable to probe the fluctuations at the edge is the position of the fermion which is {\it the farthest} from the center of the trap \cite{DDMS15,DLMS17}, hence the connection to EVS.  And since the positions of the fermions are typically strongly correlated, this is a quite hard problem. 

In fact, in dimension $d=1$ and zero temperature ($T=0$), analytical progress can be achieved in some cases thanks to a connection between the positions of the fermions and the eigenvalues of a random matrix. Let us illustrate this on the case of $N$ non-interacting spin-less fermions in a harmonic trap $V(x) = m \omega^2 x^2/2$. In this case, one can show that at $T=0$ the joint PDF of the positions $P_{\rm joint}(x_1, \cdots, x_N)$, given by the modulus square of the many-body ground state wave function, reads (see e.g. \cite{MMSV14} as well as Appendix \ref{sec:app_fermions})
\bea\label{PDF_OH}
P_{\rm joint}(x_1, \cdots, x_N) = \frac{1}{Z_N} \exp{\left(-\alpha^2\sum_{i=1}^N  x_i^2 \right)} \prod_{i<j} (x_i - x_j)^2 ~,
 \eea 
where $\alpha = \sqrt{m \omega/\hbar}$ and $Z_N$ is a normalization constant. By comparing this expression (\ref{PDF_OH}) with the joint law of the eigenvalues discussed above in Eq. (\ref{joint_GUE}) we see that the $x_i$'s behave (upto an $N$-dependent rescaling) like the eigenvalues of the GUE, i.e. $\beta = 2$. In particular, the position $x_{\max}(T=0)$ of the fermion which is the farthest from the center of the trap, i.e. the rightmost fermion, has the same statistics as the largest eigenvalue $\lambda_{\max}$ of a GUE matrix studied before \cite{DDMS15}. An immediate consequence is that the distribution of $x_{\max}(T=0)$, properly shifted and scaled, converges to the TW-distribution for the GUE, i.e. ${\cal F}'_2(x)$. Therefore, trapped fermions are probably the 
simplest physical system where the TW distribution could be observed. Interestingly, this behavior at the edge of a trapped Fermi gas is not restricted to the harmonic oscillator but actually holds (in the limit of large $N$) for a much wider class of smooth potentials of the form $V(x) \sim x^p$ with $p>0$ \cite{DDMS15,DDMS16} (although the one-to-one correspondence with the GUE for any finite $N$ only holds for the harmonic oscillator). Recently, it was argued that this also holds in the presence of (not too strong) interactions \cite{Ste19}. This edge behavior however gets modified if one considers non-smooth potentials, like a simple hard box or singular potentials. In this case, the EVS of the Fermi gas at $T=0$ is governed by the hard-edge universality class of RMT \cite{LLMS17,LLMS18}. Note also that in some cases, the large deviation of $x_{\max}(T=0)$ has also been studied \cite{LLMS17} and they show a qualitatively similar behavior as the large deviations of $\lambda_{\max}$ discussed in the previous section [see Eq. (\ref{summary})]. We refer the reader to the recent review \cite{DLMS19} for a more detailed description of these connections between one-dimensional Fermi systems at $T=0$ and RMT.  

What about the effects of finite temperature $T>0$ on the EVS of non-interacting one-dimensional Fermi systems? At very high temperature, i.e. $T \gg E_F$ where $E_F$ is the Fermi energy of the system, one expects that the quantum effects become negligible such that the fermions simply behave as classical independent particles. The corresponding EVS for this system is thus the one of IID random variables discussed above. Therefore, for smooth potentials like $V(x) \sim |x|^p$ the distribution of $x_{\max}(T \to \infty)$, appropriately shifted and scaled, converges to a Gumbel distribution while it converges to a Weibull distribution for a hard box potential. As $T$ is varied, one thus expects a crossover between EVS for strongly correlated systems (like the TW distribution in the case of smooth potentials) as $T \to 0$ to the EVS of IID variables at high temperature. It turns out that the complete 
study of the crossover in the full temperature range from $T \to 0$ to $T \to \infty$ is highly non-trivial but it was partly studied 
in Ref. \cite{DDMS15,DDMS16} for smooth potentials. In particular, in the first case, the crossover is governed by a finite-temperature generalisation of the TW distribution which, unexpectedly, arises also in exact solution of the KPZ equation in droplet geometry at finite time~\cite{DDMS15}.       

The statistics of the maximal radial distance $r_{\max}(T)$ of the fermions from the trap center was also studied in dimension $d>1$, both for smooth potentials like the harmonic well \cite{DLMS17}, and the hard box (spherical) potential \cite{LLMS17,LLMS18}. In both cases, although the positions of the fermions are strongly correlated at $T=0$, it was found that, for large $N$, the statistics of $r_{\max}(T=0)$ is given by the EVS of IID random variables, i.e. Gumbel for smooth potentials and Weibull for the spherical hard box. In Ref. \cite{DLMS17} the ``decorrelating'' mechanism behind this was elucidated. The 
main idea is to decompose the many-body ground state wave function in radial and angular sectors. For each angular quantum number, one effectively obtains a one-dimensional problem for a certain number of non-interacting fermions, in an effective quantum potential characterized by the angular quantum number. In each of these angular sectors, the distribution of the farthest fermion position is non trivial. However, for spherically symmetric potentials, 
the different angular sectors decouple and effectively one has to look at the maximum of a collection of independent, but non-identically distributed random variables. This mechanism eventually leads to the same Gumbel or Weibull law, as for the IID case, even at zero temperature. Of course, as in the $1d$ case one also expects that at $T \to \infty$ one also recovers the EVS for IID random variables, but in this case the mechanism is quite different and the study of the description of the full crossover between these two IID regimes (as $T\to 0$ and $T \to \infty$) remains an open question. Another interesting feature of the case $d>1$ concerns the large deviations of $r_{\max}(T)$. Indeed, it was shown in \cite{LLMS17,LLMS18} that, at variance with the standard scenario (\ref{summary}) found for instance for the $\beta$-Gaussian ensembles, there exists generically an {\it intermediate regime} of fluctuations between the typical and the left large deviation regime. This intermediate regime, which is actually not restricted to EVS but also concerns other observables like the full counting statistics \cite{Ber19}, seems to be a rather generic feature of higher dimensional determinantal processes, including in particular (complex) Ginibre random matrices and two-dimensional Coulomb gas \cite{LGMS18}.

\subsection{Other extreme value problems}


\subsubsection{Density of near-extremes}

While the statistics of the value of the maximum is important and has interesting limiting large $N$ behavior, as discussed extensively in this review, an equally important issue concerns the near-extreme events \cite{Sanjib}, i.e., how many events occur with their values near the extreme? In other words, the issue is whether the global maximum (or minimum) value is very far from others ({\it is it lonely at the top?}), or whether there are many other events whose values are close to the maximum (or minimum) value. This issue of the crowding of near-extreme events arises in many problems. For instance, in disordered systems, the zero temperature properties are completely governed by the ground-state, i.e. the state with {\it minimum} energy. For any finite non-zero temperature, the properties of the system are governed not just by the ground-state but the low-energy excitation states, which are just above the ground-state. Therefore, it is very important to known how many such excited states are there, above the ground-state, in an energy band $[E, E+dE]$. Such questions arise also in climate science, or in economy. For instance, for an insurance company it is very important to safeguard itself against excessively large claims. It is equally or may be more important to guard itself from an unexpectedly high number of them. Hence, it is important to know the number of insurance claims that are close to the maximal claim. In addition, in many optimization problems finding the exact optimal solution is extremely hard and the only practical solutions available are the near-optimal ones \cite{DLM05}. How many such solutions are there, whose costs are close to the cost of the optimal solution?  
In all these situations, prior knowledge about the statistics of the ``crowding'' of the solutions near the optimal one is very much desirable.

Consider a time series $x_1, x_2, \cdots, x_N$ of $N$ continuous entries, which may or may not be correlated. 
Let $x_{\max} = \max \{x_1, x_2, \cdots, x_N \}$ denote the unique maximum. A natural way to quantify the ``crowding'' near the maximum is to consider
the density w.r.t. the maximum \cite{Sanjib}
\begin{eqnarray}
\rho(r,N) = \frac{1}{N} \sum_{x_i \neq x_{\max}}^{N-1} \delta(r-(x_{\max} - x_i)) \label{def_dos} \;.
\end{eqnarray}  
The quantity $\rho(r,N)\, dr$ counts the fraction of entries with values between $r$ and $r+dr$ below the maximum, in a given realisation
of the time series. Evidently, it is a random variable, which fluctuates from realisation to realisation of the time series. Note that this is different
from the standard density of events, since $x_{\max}$ itself is a random variable that fluctuates from sample to sample. An analogous quantity can 
be defined for a continuous time stochastic process $x(\tau)$, for $\tau \in [0,t]$
\begin{eqnarray}
\rho(r,t) = \frac{1}{t} \int_0^t \delta(r-(x_{\max}(t) - x(\tau))) \, d\tau~, \label{def_dos_cont}
\end{eqnarray}
where $x_{\max}(t) = \max_{0 \leq \tau \leq t} x(\tau)$ is the global maximum of the process in $[0,t]$. The average of this quantity, either in Eq. (\ref{def_dos}) or (\ref{def_dos_cont}) has been computed exactly (i) for a time series with IID entries, each drawn from a common parent distribution $p(x)$ \cite{Sanjib}, (ii) for a Brownian motion over a fixed time interval $[0,t]$ \cite{PCMS13,PCMS15} and (iii) when the $x_i$'s represent the $N$ real eigenvalues of an $N \times N$ complex Hermitian matrix with Gaussian entries (the so called Gaussian Unitary Ensemble) \cite{PS14,PS15}. In the two latter cases, the entries of the time series are strongly correlated.

In case (i) of IID random variables, it turns out that the average of the density in Eq. (\ref{def_dos}) has been computed for all $N$ \cite{Sanjib}. In particular, for large $N$, interesting scaling behaviours emerge with universal scaling functions, depending on whether the tail of $p(x)$ for large $x$ decays faster or slower than an exponential, with a nontrivial marginal form for the exponential tail. For the Brownian motion case (ii), not just the average but also the full distribution of $\rho(r,t)$ has been computed \cite{PCMS13,PCMS15}. For the GUE matrix (iii), the average has been computed for large $N$ and its large $N$ behaviour has found application in the computation of the average susceptibility in the random $p$-spin spherical model of spin-glasses \cite{FPA15}. We do not provide further details and the interested reader may consult the original references cited above.

\subsubsection{The time at which the maximum/minimum is reached} \label{sec:tM}

An interesting observable associated to the statistics of extremes is the time at which the extreme occurs (see Fig.~\ref{Fig:BM}). For example, in a typical time series, like the price of a stock or the rainfall data, while the actual value of the extreme is important, as discussed in this article, an equally relevant question is: when does it occur? For example, in financial markets, one would naturally like to know when the price of a stock becomes maximum (that is the ideal time to sell the stock) or minimum (when one would like ideally to buy this stock). Given a stochastic process $x(\tau)$ of duration $t$, let $t_{\max}$ denote the random variable such that $x(t_{\max})$ is bigger than all other values of the process in the interval $\tau \in [0,t]$. Mathematically, this is sometimes denoted as the function ``argmax'', i.e., $t_{\max} = {\rm argmax} [x(\tau)]$, for $\tau \in, [0,t]$. Clearly, $t_{\max}$ is a random variable, with range in $[0,t]$. A natural question then is: given $t$, what is the PDF of $t_{\max}$? Note that the same question can be asked for a discrete time series $\{x_1, x_2, \cdots, x_N\}$ of $N$ entries. In this case, this random variable will be denoted by $n_{\max} = {\rm argmax}\{x_1, x_2, \cdots, x_N\}$ which is clearly discrete and $1 \leq n_{\max} \leq N$. 

We start with the simple example of a discrete time series $\{x_1, x_2, \cdots, x_N \}$ of $N$ IID entries, each drawn from a PDF $p(x)$.
As the maximum (or the minimum) can be any one of the entries with equal probability $1/N$, it follows that the distribution of $n_{\max}$ is just the uniform distribution 
\begin{eqnarray}\label{nmax_iid}
P(n_{\max}  |N) = \frac{1}{N} \quad, \quad n_{\max}=1,2, \cdots, N \;.
\end{eqnarray} 
Thus this result is completely universal, i.e., independent of the PDF $p(x)$. While in this IID case the result is very simple and universal, one naturally wonders what is the distribution of $n_{\max}$ when the entries of the time series are correlated, in particular when the correlations are strong, in the sense discussed before.

The simplest strongly correlated stochastic process is the continuous-time Brownian motion $x(\tau)$ [see Eq. (\ref{def_BM})] over the time interval $\tau \in [0,t]$. In this case, the PDF of $t_{\max}$, denoted by $P(t_{\max}|t)$, is a continuous distribution and was computed a long time ago by P. L\'evy \cite{Lev40}. It  has a very simple expression 
\begin{eqnarray}\label{PDF_arcsine}
P(t_{\max}|t) = \frac{1}{\pi \,\sqrt{t_{\max}(t-t_{\max})}} \quad, \quad 0 \leq t_{\max} \leq t \;.
\end{eqnarray}
The corresponding cumulative distribution reads
\begin{eqnarray}\label{arcsine}
{\rm Prob.}(t_{\max} \leq T) = \int_{0}^T \, P(t_{\max}|t) \ dt_{\max}  = \frac{2}{\pi} {\rm arcsin}\left( \sqrt{\frac{T}{t}}\right) \;,
\end{eqnarray} 
which is known as the celebrated ``arcsine law'' of L\'evy. Note that the PDF (\ref{PDF_arcsine}) diverges with a square-root singularity at the two
edges $t_{\max} = 0$ and $t_{\max} = t$. Since the Brownian motion starts at the origin, the event ``$t_{\max} = 0$'' corresponds to trajectories that start at the origin and stay negative over the full interval $[0,t]$. In contrast, the event ``$t_{\max} = t$'' corresponds to trajectories where the maximum occurs at the end of the interval $[0,t]$. Also the minimum of the curve occurs at $t_{\max} = t/2$, which is also the average value of $t_{\max}$. Thus instead of having a peak at its average value, the PDF has peaks at the two ends. This indicates that the typical $t_{\max}$ is either $0$ or $t$, while the average is $t/2$.   

One can also ask the same question for a discrete-time random walk defined before in Eq. (\ref{eq:RW}) where 
the position evolves according to $x_{n} = x_{n-1} + \eta_n $, starting form $x_0 = 0$ and where the jump lengths at each step are drawn independently from a symmetric and continuous distribution $f(\eta)$. Note that this includes long-range jumps such as L\'evy flights, for which $f(\eta)$ has a heavy tail $f(\eta) \sim 1/|\eta|^{1+\mu}$ as $|\eta| \to \infty$ with $0< \mu < 2$. In this case, the distribution of $n_{\max}$ turns out to be completely universal, i.e., independent of the jump density $f(\eta)$ and is given by \cite{SA53}
\begin{eqnarray}\label{nmax}
P(n_{\max} | N) = {2n_{\max} \choose  n_{\max}} {2(N-n_{\max}) \choose (N-n_{\max})} 2^{-2N} \;.
\end{eqnarray}   
This is a consequence of the celebrated Sparre Andersen theorem. For a simple derivation of this result (\ref{nmax}), we refer the reader to Ref. \cite{satya_review10}. In the limit of large $m$ and $n$ (keeping the ratio $m/n$ fixed), one gets from Eq. (\ref{nmax}), using Stirling's formula 
\begin{eqnarray}\label{nmax_asympt}
P(n_{\max} | N) \approx \frac{1}{\pi \sqrt{n_{\max}(N-n_{\max})}} \;.
\end{eqnarray}
Naively, this resembles the result in Eq. (\ref{PDF_arcsine}) for the Brownian motion, but this is a bit deceptive. Note indeed that this formula (\ref{nmax_asympt}) holds even for L\'evy flights, which could not have been derived from the continuous-time Brownian motion result.

The distribution of $t_{\max}$ has also been computed, using path-integral methods, for a variety of constrained Brownian motions, such as Brownian bridge, Brownian excursion, Brownian meander, reflected Brownian motion \cite{RM07, MRKY08}, Brownian motion with a drift \cite{MB08}, etc. These results have also been obtained using real-space renormalization group methods, which also allowed exact computation for the distribution of $t_{\max}$ of other stochastic processes such as the continuous-time random walk (CTRW) and the Bessel process that describes the radial component of the Brownian motion in $d$ dimensions \cite{SL10}. Brownian motion, or random walks, are Markov processes. Recently, the distribution of $t_{\max}$ has been computed exactly for some non-Markov processes. This includes the random acceleration process, i.e. $d^2x/dt^2 = \eta(t)$ where $\eta(t)$ is a Gaussian white noise \cite{MRZ110}. Another example is the fractional Brownian motion with Hurst index $H \in [0,1]$. The Brownian motion corresponds to $H=1/2$ and the distribution of $t_{\max}$ has recently been computed~\cite{DW16,DW16b,DRW17,SDW18}, using a perturbation theory around the Markov limit $H = 1/2$ developed in Ref. \cite{WMR11}. 

There have been many interesting applications of $P(t_{\max}|t)$. For example, the distribution of $t_{\max}$ appears in a class of one-dimensional diffusion problems in a disordered potential, such as in the Sinai model. In fact, replacing $t_{\max}$ by $x$ and $t$ by $L$, the PDF $P(t_{\max} = x|t=L)$ describes the equilibrium disordered average probability density of the position $x$ of a particle in a disordered potential inside a box of size $L$ \cite{MRZ10}. Another application concerns the convex hull of Brownian motion in two-dimensions. It has been shown that, to compute the mean area of the convex hull of a two-dimensional stochastic process, the knowledge of $P(t_{\max}|t)$ of the one-dimensional component of the $2d$ motion is needed \cite{RMC09,MCR10,RMR11}. 

So far, we have been discussing the distribution of $t_{\max}$ for a single particle problem. However, it is also interesting to ask how this
distribution is affected in the presence of more than one particles, independent or interacting. For $n$ independent Brownian motions, the 
distribution of $t_{\max}$ can be computed exactly and this result has been used to compute the mean area of the convex hull of $n$ independent   
Brownian motions in two-dimensions \cite{RMC09,MCR10}. An example of an interacting particle system where the distribution of $t_{\max}$
can be computed concerns $n$ vicious walkers in one-dimension \cite{RS11}. This result has implication for $1+1$-dimensional growth models belonging
to the Kardar-Parisi-Zhang (KPZ) universality class \cite{BLS12,FQR13}.

\subsubsection{Records} 

We end this review by discussing briefly another topic which is intimately related to extreme statistics, namely the statistics of records which have found 
applications in finance, sports, climates, ecology, disordered systems, all the way to evolutionary biology. Consider any generic time series of $N$ entries $X_1, X_2, \cdots, X_N$ where $X_i$ may represent the daily temperature in a given place, the price of a stock or the yearly average water level in a river. A record happens at step $k$ if the $k$-th entry exceeds all previous entries. Questions related to records are obviously intimately connected to EVS. For instance the actual record value at step $k$ is precisely 
the maximal value of the entries after $k$ steps, which we have discussed in this review. In fact, as for the case of EVS, the statistics of records for
IID sequences is perfectly well understood (see e. g. Ref. \cite{Nev}). However, much less is known for strongly correlated time series. During the last years, important progress has been achieved in the statistics of records of random walks and L\'evy flights (see e.g. \cite{MZ08}), which, also for records, serve as a very useful laboratory to test the effects of strong correlations and we refer the reader to the recent survey \cite{GMS17} for a detailed account of these results on record statistics for correlated random variables.

\section{Summary and Conclusion}
To conclude, we have made a brief overview 
on the subject of extreme value statistics. We have seen that 
for uncorrelated or weakly correlated random variables, one
has a fairly good understanding of the distribution of extremum
and their limiting laws:
there exists essentially three limiting classes named {Fr\'{e}chet,~Gumbel,~Weibull} respectively.
On the other hand, there are very few exact results known for the 
{\em strongly} correlated random variables. In this review, 
we have discussed a few of them, but were not able to cover all
of them. Most of the theoretical efforts are focussed in finding
more and more exactly solvables cases which may shed some
light on the issue of the universality classes of EVS for
strongly correlated random variables. Identifying the
universality classes (if they exist) for strongly correlated
random variables is thus a very challenging and outstanding 
open problem.

Apart from the theoretical issue of identifying 
the universality classes of the extreme value 
distributions for strongly correlated variables, the extreme statistics of
such variables also have found a large number of applications in a variety
of problems arising in physics, mathematics, computer science, biology all the way
to climatology or finance. In this review we have discussed very briefly some of these applications
but we were not able to give an exhaustive account of these applications. For example, the
largest eigenvalue of a random matrix, the limiting Tracy-Widom distributions, and also
its large deviation tails, have found a wide variety of applications and have also been measured experimentally (for 
a brief review see \cite{SM13} and the reader may also consult a couple of popular articles on the subject \cite{buchanan, wolchover}).

Going beyond the statistics of the maximum or the minimum value, 
there are other important and relevant questions related to extremes. For
example, in this review, we have briefly touched upon some of these questions such as 
the density of near-extreme events, the statistics of the time at which the extreme occurs 
and the statistics of records. Evidently, these questions also have several applications. 
Another interesting topic that is of much current interest concerns the statistics of the number
of maxima, minima or the saddle points of a random high-dimensional landscape. This is of interest
in a variety of fields \cite{Wales04}, including population dynamics \cite{May1972,MayBook}, models of evolutionary biology \cite{NK11,PK16}, spin-glasses \cite{GD88,BDG01,CC05,MM06,BD07,SMB08,FN12,ABC13,Fyo15,RBBC19}, neural
networks \cite{WT13} as well as in landscape based string theory \cite{string1,string2} and cosmology \cite{AE06}. Recently, there has been a surge of interest in these questions in the context of deep learning 
and artificial intelligence \cite{MM_book09,KMSSZ12,deep_2,deep_1}. Discussing these issues is beyond the scope of this review and we may refer the reader to recent interesting reviews mentioned above.

In this short review, we have tried to give an account of the current and past research in the rapidly evolving field of extreme value statistics, in particular
focusing on various physical applications of EVS. This review, by no means, provides an exhaustive list of applications of EVS in other fields, such as in biology or computer science, neither it provides details of derivations. The idea is to provide general pointers to some of the recent developments in the field
in the context of statistical physics. We hope that this review will stimulate further interests and developments in this field.

\vspace*{1cm}

 {\acknowledgements}
 
 It is a pleasure to thank many of our collaborators and colleagues: G. Akemann, R. Allez, J. Baik, A. Bar, M.~Barma, M. Battilana, G. Ben Arous, E. Ben Naim, O. B\'enichou, B. Berkowitz, G. Biroli, O. Bohigas, G. Borot, J.-P.~Bouchaud, D. Boyer, A. J. Bray, J. Bun, P. Calka, T. Clayes, A. Comtet, I. Corwin, F. D. Cunden, C. Dasgupta, D. S. Dean, B. Derrida, A. Dhar, D. Dhar, C. De Bacco, B. Eynard, M. R. Evans, P. J. Forrester, Y. V. Fyodorov, T. Gautier, C. Godr\`eche, A. Grabsch, D. Grebenkov, J. Grela, S. Gupta, A. Hartmann, H. J. Hilhorst, M. Katori, E. Katzav, M. J. Kearney, A. Krajenbrink, P. L. Krapivsky, J. Krug, A. Kundu, B. Lacroix-A-Chez-Toine, A. Lakshminarayan, P.~Leboeuf, K. Liechty, P. Le Doussal, J.-M. Luck, K. Mallick, M. Mari\~no, R. Marino, O. C. Martin, B. Meerson, M.~M\'ezard, C.~Monthus, F.~Mori, Ph.~Mounaix, D.~Mukamel, C. Nadal, S. K. Nechaev, N. O'Connell, H. Orland, G.~Oshanin, M. Potters, I. Perez-Castillo, A. Perret, J. Pitman, Z. Racz, J. Rambeau, K. Ramola, J. Randon-Furling, A. Rosso, S. Redner, S.~Reuveni, S. Sabhapandit, A. Scardicchio, H. Schawe, C. Sire, P. Sollich, H. Spohn, C. Texier, M. Tierz, S. Tomsovic, C. A. Tracy, D. Villamaina, M. Vergassola, P. Vivo, G. Wergen, S. R. Wadia, K. J. Wiese, M.~Yor, O.~Zeitouni, R. K. P. Zia, R. M. Ziff, A. Zoia. S.N.M. and G. S. acknowledge support from ANR grant ANR-17-CE30-0027-01 RaMaTraF.

\vspace*{1cm}

\appendix

\section{Some details about the Tracy-Widom distribution}\label{Sec:TW}

In this Appendix, we give some details about the Tracy-Widom distributions (and its generalizations), which describe the fluctuations of the largest eigenvalue
in the Gaussian $\beta$-ensembles, i.e. matrix models whose eigenvalues are distributed according to Eq. (\ref{joint_GUE}). As stated in the text, the largest eigenvalue $\lmax$ in these ensembles behaves, in the large $N$ limit, as
\begin{eqnarray}\label{chi_beta}
\lmax = \sqrt{2} + \frac{1}{\sqrt{2}}\, N^{-2/3}\, \chi_\beta \;,
\end{eqnarray}
where $\chi_\beta$ is an $N$-independent random variable. Its cumulative distribution function (CDF), ${\cal F}_{\beta}(x)= {\rm Prob.}[\chi_\beta\le x]$, 
is known as the $\beta$-Tracy-Widom (TW) distribution. 

\vspace*{0.4cm}
\noindent{\bf The case $\beta = 2$}. This case corresponds to the so-called Gaussian Unitary Ensemble (GUE) of RMT. For this special value of $\beta$, the eigenvalues form a determinantal point process and the TW distribution ${\cal F}_2$ can be written in terms of Fredholm determinant as
\begin{eqnarray} \label{Fredholm_Airy}
{\cal F}_2(s) = {\rm Det}({\mathbb I} - P_s K_{\rm Ai} P_s) \;, 
\end{eqnarray}
where $K_{\rm Ai}(x,y)$ is the so-called Airy kernel, given by
\begin{eqnarray}\label{Airy_K}
K_{\rm Ai}(x,y) = \frac{{\rm Ai}(x) {\rm Ai}'(y) - {\rm Ai}'(y) {\rm Ai}(x)}{x-y} = \int_0^\infty {\rm Ai}(x+u) {\rm Ai}(y+u) \, du \;,
\end{eqnarray}
where ${\rm Ai}$ and ${\rm Ai}'$ denote the Airy function and its derivative respectively. We recall that a Fredholm determinant, as in Eq. (\ref{Fredholm_Airy}), is conveniently expressed by the trace expansion using ${\rm Det}({\mathbb I} - P_s K_{\rm Ai} P_s) = \exp(-\sum_{p\geq 1} {\rm Tr}(P_s K_{\rm Ai} P_s)^p/p)$, where $P_s$ denotes the projector on the interval $[s, + \infty)$, i.e. $[P_s K_{\rm Ai} P_s] (x,y)= \theta(x-s) K_{\rm Ai}(x,y) \theta(y-s)$ where $\theta(z)$ is the Heaviside step function. In Ref. \cite{TW94}, Tracy and Widom showed that this Fredholm determinant can be written in terms of a special solution of the Painlev\'e II equation (the so-called Hastings-McLeod solution), denoted $q(s)$, which satisfies 
\begin{eqnarray} \label{PII}
q''(s) = 2 q^3(s) + s q(s) \;, \; q(s) 
\sim {\rm Ai}(s) \;, \, s \to \infty \;.
\end{eqnarray}
In terms of this function (\ref{PII}), the TW distribution for $\beta=2$, ${\cal F}_2$ admits a rather simple expression, namely
\begin{eqnarray}\label{TW_F2}
{\cal F}_2(s) = \exp{\left[-\int_s^\infty  (s'-s) q^2(s') \, ds'\right]} \;.
\end{eqnarray}
We refer the reader to Ref. \cite{NM11} for a derivation of this expression (\ref{TW_F2}) of the TW distribution for $\beta=2$, without using the representation in terms of a Fredholm determinant (\ref{Fredholm_Airy}) but using instead (non-classical) orthogonal polynomials.

\vspace*{0.4cm}
\noindent{\bf The case $\beta = 1, 4$}. In this case, the computation of ${\cal F}_1(s)$ and ${\cal F}_4(s)$ are more complicated because the eigenvalues form a Pfaffian point process -- which is a slightly more complex structure than a determinantal point process. Hence, ${\cal F}_{1}(s)$ and ${\cal F}_4(s)$ are then given by Fredholm Pfaffians. However, Tracy and Widom were able to reduce these Fredholm Pfaffians to Fredholm determinants \cite{TW96} and from that, they could express ${\cal F}_1(s)$ and ${\cal F}_4(s)$ also in terms of the solution of the Painlev\'e equation (\ref{PII}). Namely, they obtained \cite{TW96}
\begin{eqnarray}\label{TW_124}
&&{\cal F}_1(x) = \exp{\left[-\frac{1}{2} \int_x^\infty \left[ (s-x) q^2(s) + q(s) \right] \, ds\right]} \;, \\
&&{\cal F}_4(2^{-\frac{2}{3}}x) = \exp{\left[-\frac{1}{2}  \int_x^\infty  (s-x) q^2(s) \, ds\right]} \cosh{\left[\frac{1}{2} \int_x^\infty q(s)\, ds \right]} \;, \nonumber
\end{eqnarray}
where $q(s)$ is given in Eq. (\ref{PII}). 

\vspace*{0.4cm}
\noindent{\bf Other values of $\beta \neq 1, 2, 4$}. For other values of $\beta$, there is no such explicit expression for ${\cal F}_\beta(s)$ similar to (\ref{TW_F2}) and (\ref{TW_124}) [see however the case $\beta=6$ Ref. \cite{Rum16,GIKM16} for which ${\cal F}_6(s)$ can also be expressed in terms of Painlev\'e transcendents, albeit in a more complicated way than Eqs. (\ref{TW_F2})-(\ref{TW_124})]. Nonetheless, for any generic value $\beta >0$, the random variable $\chi_\beta$ in Eq. (\ref{chi_beta}) admits an interesting representation. Indeed, it can be shown that $\chi_\beta$ describes 
the fluctuations of the ground state of the following one-dimensional 
Schr\"odinger operator, called the ``stochastic Airy operator'' 
\cite{ES07, RRV11}
\begin{eqnarray}\label{stoc_Airy}
{\cal H}_{\beta} = -\frac{d^2}{dx^2} + x + \frac{2}{\sqrt{\beta}} \eta(x) \;,
\end{eqnarray}
where $\eta(x)$ is Gaussian white noise, of zero mean and with delta 
correlations, $\langle{\eta(x) \eta(x')\rangle} = \delta(x-x')$.

\vspace*{0.4cm}
\noindent{\bf Precise asymptotic tails}. We end up this section by indicating the precise asymptotic tail of ${\cal F}_\beta(s)$, beyond the leading order given in Eq. (\ref{tail_TW}). The most precise asymptotic expansions, both for the right and left tail, of ${\cal F}_\beta(s)$ have been obtained in the physics literature by using the so-called loop-equations. For the right tail, i.e. in the limit $s\to \infty$, the asymptotic behavior of ${\cal F}_\beta(s)$  reads \cite{BN12}
\begin{eqnarray} \label{right_tail}
1 - {\cal F}_\beta(s) = \frac{\Gamma(\frac{\beta}{2})}{2 \pi\,(4 \beta)^{\beta/2}} \, s^{-\frac{3\beta}{4}}\, e^{- \frac{2\beta}{3} s^{3/2}} \exp{\left[\sum_{m\geq 1} \frac{\beta}{2} R_m\left( \frac{2}{\beta}\right) \, s^{- \frac{3m}{2}}\right]}
\end{eqnarray}   
where $R_m(z)$ are polynomials with rational coefficients, of degree at most $m+1$ and which can be computed recursively \cite{BN12}. The two first leading terms in (\ref{right_tail}) are in agreement with the rigorous result \cite{DV13} $1 - {\cal F}_\beta(s) = s^{-\frac{3\beta}{4} + o((\ln s)^{-1/2})}\, e^{- \frac{2\beta}{3} s^{3/2}}$ obtained in Ref. \cite{DV13} using the representation of $\chi_\beta$ in terms of the stochastic Airy operator~(\ref{stoc_Airy}). 

The left tail of ${\cal F}_\beta(s)$, for $s \to -\infty$, is also known with high precision and it reads
\begin{eqnarray} \label{left_tail}
{\cal F}_\beta(s) = \exp{\left[-\beta \frac{|s|^3}{24} + \frac{\sqrt{2}}{3}(\beta/2-1) |s|^{3/2} + \frac{1}{8} (\beta/2+2/\beta-3) \, \ln |s| + c_\beta + O(|s|^{-3/2}) \right]} \;,
\end{eqnarray}
where $c_\beta$ has a complicated, though explicit, expression \cite{BEMN11}. This asymptotic behaviour (\ref{left_tail}) has been proven rigorously for $\beta = 2$ in Ref. \cite{BBD08} and for $\beta = 1,4$ in \cite{DIK08}.

\section{From noninteracting fermions in a harmonic trap at $T=0$ to random matrices}\label{sec:app_fermions}

In this section, we provide the derivation of the joint PDF 
of the positions, given in Eq. (\ref{PDF_OH}), of $N$ noninteracting fermions in a harmonic potential $V(x) = \frac{1}{2} m\omega^2 x^2$ in their ground state. The system is described by the $N$-body Hamiltonian
\begin{eqnarray} \label{H_N}
{\cal H}_N = \sum_{j=1}^N \hat h(\hat x_j, \hat p_j) \; \quad , \quad \; \hat h(\hat x, \hat p) = \frac{\hat p^2}{2m} + \frac{1}{2} m\omega^2 \hat x^2 \;.
\end{eqnarray}
The single particle eigenstates $\phi_k(x)$, such that $\hat h \phi_k(x) = \epsilon_k \phi_k(x)$, can be obtained exactly 
\begin{eqnarray}\label{hermite_fct}
\phi_k(x) = \left[ \frac{\alpha}{\sqrt{\pi} 2^{k-1} (k-1)!}\right]^{1/2} \, e^{-\frac{\alpha^2\,x^2}{2}} H_{k-1}(\alpha \,x) \;,
\end{eqnarray} 
where $k=1,2, \cdots$ and $H_k(z)$ is the $k$-th Hermite polynomial of degree $k$ and $\alpha = \sqrt{m \omega/\hbar}$ is the characteristic length scale of the trap. The associated single-particle energy levels are given by $\epsilon_k = (k-1/2)\, \hbar \omega$. The ground-state of the $N$-body system corresponds to filling up the $N$ first single-particle energy levels with
one fermion per level (as dictated by the Pauli exclusion principle). Correspondingly, the $N$-body ground-state wave-function 
is given by the Slater determinant
\bea\label{Slater_1d}
\Psi_0(x_1, \cdots, x_N) = \frac{1}{\sqrt{N!}} \, \det_{1\leq \,j,\,k\, \leq N}\phi_{k}(x_j)  \;,
\eea
with the associated energy $E_0 = \sum_{k=1}^N \epsilon_k$. The quantum probability density function (PDF) is then given by
\bea\label{jpdf}
P_{\rm joint}(x_1, \cdots, x_N) = |\Psi_0(x_1, \cdots, x_N)|^2 = \frac{1}{N!} \left|\det_{1\leq \,j,\,k\, \leq N} \phi_{k}(x_j) \right|^2 \;.
\eea
This joint PDF is normalised  and encodes the quantum fluctuations of the Fermi gas. In the Slater determinant, the Gaussian factors in Eq. (\ref{hermite_fct}) come out of the determinant, leaving us to compute the determinant of a matrix consisting of Hermite polynomials. The Hermite polynomials $H_0(z), H_1(z), \cdots, H_{N-1}(z)$ provide a basis for polynomials of degree $N-1$ and by  manipulating the rows and columns, the determinant can  be reduced to a Vandermonde determinant. Hence, we can evaluate the Slater determinant in (\ref{jpdf}) explicitly to obtain the formula in Eq. (\ref{PDF_OH}).

\end{document}